\documentclass[usenatbib,preprint]{mn2e}

\usepackage{graphicx}
\usepackage {natbib,aas_macros}
\usepackage {mediabb}
\usepackage {bm}
\usepackage {amsmath,amssymb}
\usepackage {setspace}

\newcommand{\msun}{\thinspace M_\odot} 
\newcommand{\gcm}{~{\rm g~cm}^{-3} } 
\newcommand{\Jang}{\mathbf{J_{\rm ang}}}

\newcommand{\magB}{\mathbf{B}}

\newcommand{\cmcms}{~{\rm cm}^2 ~{\rm s}^{-1} }

\title{Does  misalignment between magnetic field and angular momentum 
enhance or suppress circumstellar disk formation?}

\author[Tsukamoto et al]{
Y. Tsukamoto$^{1}$,   S. Okuzumi$^{2}$, K. Iwasaki$^{3}$, M. N. Machida$^{4}$, and  S. Inutsuka$^{5}$ \\
$^1$Graduate Schools of Science and Engineering, Kagoshima University, Kagoshima, Japan  \\
$^2$Department of Earth and Planetary Sciences, Tokyo Institute of Technology, Tokyo, Japan \\
$^3$Department of Earth and Space Science, Osaka University, Osaka, Japan \\
$^4$Department of Earth and Planetary Sciences, Kyushu University, Fukuoka, Japan \\
$^5$Department of Physics, Nagoya University, Aichi, Japan  \\
}

\begin{document}
\maketitle

\begin{abstract}
The effect of misalignment between the magnetic field $\magB$ 
and the angular momentum $\Jang$
of molecular cloud cores on the angular momentum evolution during 
the gravitational collapse is investigated by ideal and non-ideal MHD simulations.
For the non-ideal effect, we consider the ohmic and ambipolar diffusion.
Previous studies that considered the misalignment 
reported qualitatively contradicting results. Magnetic braking was reported as
being either strengthened or weakened by misalignment in different studies.
We conducted simulations of 
cloud-core collapse by varying the stability parameter $\alpha$
(the ratio of the thermal to gravitational energy of the core)
with and without including magnetic diffusion
The non-ideal MHD simulations
show the central angular momentum of the core with 
$\theta=0^\circ$ ($\Jang \parallel \magB$) 
being always greater than that with $\theta=90^\circ$ ($\Jang \perp \magB$),
independently of $\alpha$, meaning that 
circumstellar disks form more easily form in a core with $\theta=0^\circ$.
The ideal MHD simulations, in contrast, show the 
the central angular momentum of the core with $\theta=90^\circ$ being greater than with $\theta=0^\circ$ for small $\alpha$, and is smaller for large $\alpha$.
Inspection of the angular momentum evolution of the fluid elements
reveals three mechanisms contributing to
the evolution of the  angular momentum:
(i) magnetic braking in the isothermal collapse phase,
(ii) selective accretion of the rapidly (for $\theta=90^\circ$ ) 
or slowly (for $\theta=0^\circ$) 
rotating fluid elements to the central region, and
(iii) magnetic braking in the first-core and the disk.
The difference between the ideal and non-ideal 
simulations arises from the different efficiencies of (iii).
\end{abstract}


\section{Introduction}
\label{intro}
The role of magnetic field in the gravitational collapse 
of a molecular cloud core
has generated much discussion in the context of the formation and 
early evolution of circumstellar disks and of the formation of outflows and jets. 
\citep[e.g.,][]{1994ApJ...432..720B,1995ApJ...452..386B,
2003ApJ...599..363A,2008ApJ...681.1356M,2010ApJ...718L..58I,
2011PASJ...63..555M,2012A&A...541A..35D,
2012MNRAS.427.3188B,2017MNRAS.470.1026M}.
Magnetic field transfers the angular momentum from the inner 
rapidly-rotating region of a core to its outer slowly-rotating region. 
The angular momentum of the inner region therefore decreases.
This process is known as magnetic braking.
Simulations of the gravitational collapse of 
typically magnetized cloud cores have shown that magnetic braking
is efficient and has a strong impact on the formation 
and evolution of circumstellar disks 
\citep{2007Ap&SS.311...75P,2007ApJ...670.1198M,2008ApJ...681.1356M,
2010A&A...521L..56D,
2011ApJ...738..180L,2012A&A...541A..35D,2015ApJ...801..117T,2014MNRAS.438.2278M,
2015MNRAS.451..288L,2015ApJ...810L..26T,2017MNRAS.467.3324L,2017PASJ...69...95T}. 
Reviews of the protostellar collapse process and of magnetic braking can be 
found in \citet{2012PTEP.2012aA307I} and \citet{2016PASA...33...10T}, respectively.

It has been suggested that a misalignment between the magnetic field
and the angular momentum of the cloud core changes the efficiency of magnetic braking
\citep{1985A&A...142...41M, 
2004ApJ...616..266M,2006ApJ...645.1227M,
2009A&A...506L..29H,2012A&A...543A.128J,2013ApJ...774...82L,
2017MNRAS.466.1788W,2018MNRAS.473.2124G}.
However, there is qualitative disagreement between previous studies.
Using analytic calculations, \citet{1985A&A...142...41M} suggested 
that magnetic braking in the case with $\theta=90^{\circ}$ (the perpendicular configuration) is much stronger 
than with $\theta=0^{\circ}$ (parallel configuration),
where $\theta$ is the angle between the magnetic field and 
the angular momentum.
 
\citet{2004ApJ...616..266M} 
conducted the first three-dimensional simulations 
of the gravitational collapse of a molecular cloud core 
with $\magB \nparallel \Jang$
(where $\Jang$ and $\magB$ are the angular momentum and magnetic field of the core)
and found that the angular 
momentum of the central region is more efficiently removed when 
the initial magnetic field and angular momentum of the core are 
in the perpendicular configuration (i.e., $\theta=90^\circ$).
In particular, they unequivocally showed that the perpendicular component
of the angular momentum relative to the magnetic field 
is selectively removed from the central region
during the collapse of the cloud core \citep[figure 12 in][]{2004ApJ...616..266M}.
Ideal MHD simulations in \citet{2017MNRAS.466.1788W} show that
the binary formation is suppressed in a core with the perpendicular configuration 
although the binary is formed in the core with parallel configuration, 
which also implies that the magnetic braking is more efficient in the perpendicular configuration.

In contrast,
\citet{2009A&A...506L..29H} and \citet{2012A&A...543A.128J} 
reported that the efficiency of magnetic braking 
decreases as  $\theta$   
increases and is minimized in the perpendicular configuration
($\theta=90^\circ$). This
contradicts the results of \citet{2004ApJ...616..266M}.
\citet{2013ApJ...774...82L}, who also reported 
that the angular momentum of the central region is
greater for the perpendicular than for the parallel configuration.
They suggested that the transfer of angular momentum by outflows plays a key role.
All these studies adopted the ideal MHD approximation 
(while nonetheless imposing a small uniform resistivity to
resolve the numerical difficulty 
in the simulation in \citet{2013ApJ...774...82L}).
The physical process considered is almost the same in all these studies but
the results are contradictory nevertheless.

More recently,
\citet{2016A&A...587A..32M} proposed,
by incorporating the ohmic and ambipolar diffusions, that misalignment does not affect
the magnetic-braking efficiency.
The degree of ionization of the cloud core is low and 
the ideal MHD approximation is not valid in the high density region 
\citep[$\rho\gtrsim 10^{-13}\gcm$;][]
{1980PASJ...32..405U,1991ApJ...368..181N,2002ApJ...573..199N}.
Non-ideal effects (i.e., ohmic diffusion, Hall effect,
and ambipolar diffusion) should therefore be considered.
The results of \citet{2016A&A...587A..32M} suggests 
that the magnetic diffusion eliminates the effect 
of misalignment on the magnetic-braking efficiency.
Their results also imply that 
the difference between the magnetic-braking efficiency in the 
parallel and perpendicular configurations observed in
ideal MHD simulations originates in the high-density region
$\rho\gtrsim 10^{-13}\gcm$
because magnetic diffusion becomes 
dynamically important only in this region  
\citep{1990MNRAS.243..103U,1991ApJ...368..181N,2002ApJ...573..199N,
2015MNRAS.452..278T,2015ApJ...801..117T}.
We should note, however, that \citet{2016A&A...587A..32M} investigated only
the core with $\theta<40^\circ$. It is therefore unclear whether their claim also holds for
$\theta>40^\circ$ was not clear.
\citet{2017MNRAS.466.1788W} reported that
the separation of the binary formed in the core with the perpendicular configuration 
is less than with the parallel configuration. This implies
that the magnetic braking is more efficient with the perpendicular configuration
even with the non-ideal effects.

In summary, the earlier studies have reported apparently 
contradicting results, leaving the question open as to
whether magnetic braking is enhanced by, suppressed by, 
or independent of misalignment.

Resolving this discrepancy is of interest 
not only for theorists but also for observational astronomers. 
Several observations of dust polarized emission
have identified the direction of the
magnetic field on various scales in molecular clouds
\citep{2006Sci...313..812G,2013ApJ...768..159H,2014ApJ...797...74D,
2014ApJS..213...13H,2015ApJ...798L...2S,2017ApJ...842...66W,
2017ApJ...842L...9H,2017ApJ...847...92H,2018A&A...616A..56A,2018ApJ...861...65S,
2018MNRAS.477.2760M,2018arXiv180405801G,2018ApJ...855...92C,2018ApJ...859..165S,
2018ApJ...859....4K,2018arXiv180507348K}.

Furthermore, recent high-resolution observations of Class 0/I YSOs
have reported the existence of Keplerian disk with size of several 
10s to 100s AU scales \citep{2012Natur.492...83T,2012ApJ...754...52T,
2013A&A...560A.103M,2014ApJ...796...70C,2014ApJ...796..131O,2015ApJ...812...27A,
2015ApJ...812..129Y,2017ApJ...834..178Y,2017ApJ...849...56A}.
This information allows a discussion of the relative orientations 
of the magnetic field and the circumstellar disks, and of
the correlation between the disk size and magnetic-field direction.
Understanding the nature of magnetic braking in a
misaligned system is necessary to interpret such
observational results appropriately and
to discuss the importance of a magnetic field on the evolution of angular momentum.

This paper aims to resolve the apparent discrepancies between the earlier theoretical studies.
In particular, we focus on 
the dependence of magnetic braking on the gravitational stability of the 
initial cloud core, i.e., the degree of departure from
gravitational equilibrium and the mass-accretion rate onto the disk.
Following convention, the gravitational stability of the cloud core 
is parameterized by 
$\alpha\equiv E_{\rm therm}/E_{\rm grav}$,
where $E_{\rm therm}$ and $E_{\rm grav}$ are 
the thermal and gravitational energies of the initial core, respectively.
\citet{2016MNRAS.463.4246M} demonstrated
that the magnetic-braking efficiency increases as $\alpha$ increases.
However, \citet{2016MNRAS.463.4246M} only investigated the core with the
parallel configuration. On the other hand,
previous studies concerning misalignment did not carefully consider the
dependence of the result on $\alpha$.
We also conducted simulations with ohmic and ambipolar diffusions, to
compare with the results by \citet{2016A&A...587A..32M}.



The present paper is organized as follows.
We first describe the numerical methods and initial conditions
used in this study in \S \ref{method}. 
The simulation results are then shown in \S \ref{results}.
We summarize our numerical results and discuss
their implications in \S \ref{discussion}.
Finally, we present our conclusions on the impact of 
misalignment in \S \ref{conclusion}.
Appendix \ref{analytic_arg}, 
reviews the analytic discussion of the magnetic-braking timescale.
The derived timescales are used for discussions.


\section{Numerical Method and Initial Conditions}
\label{method}
\subsection{Numerical Method}
The simulations solve
non-ideal radiation magneto-hydrodynamics equations 
with self-gravity:
\begin{eqnarray}
\frac{D \mathbf{v}}{D t}&=&-\frac{1}{\rho}\left\{ \nabla
\left( P+\frac{1}{2}|\magB|^2 \right) - \nabla \cdot (\mathbf{ B B})\right\}  - \nabla \Phi,  \\
\frac{D}{D t}\left( \frac{\mathbf B}{\rho}\right) &=&\left(\frac{\mathbf B}{\rho} \cdot \nabla \right)\mathbf v    \nonumber \\ 
&-& \frac{1}{\rho} \nabla \times \left\{ \eta_O (\nabla \times \mathbf B)   \right.    \nonumber \\
&-& \left. \eta_A ((\nabla \times \mathbf B) \times \mathbf {\hat {B}}) \times \mathbf {\hat {B}}\right\},  \\
\frac{D}{D t} \left ( \frac{E_r}{\rho} \right ) &=& - \frac{\nabla \cdot \mathbf{F}_{\bm r}}{\rho} - \frac{\nabla \mathbf{v} : \mathbb
{P}_r}{\rho} +\kappa_P c (  a_r T_g^4 - E_r),  \\
\frac{D}{D t} \left ( \frac{e}{\rho} \right ) &=& - \frac{1}{\rho} \nabla \cdot \left \{ ( P+\frac{1}{2}|\magB|^2) \mathbf{v} -\mathbf{B} (\mathbf{B}\cdot\mathbf{v}) \right \} \nonumber \\
&-& \kappa_P c (  a_r T_g^4 - E_r)-\mathbf{v}\cdot\nabla \Phi  \nonumber \\
&-& \frac{1}{\rho} \nabla \cdot \left[ \left\{(\eta_O (\nabla \times \mathbf B)  \right. \right. \nonumber \\
 &-&  \left. \left. \eta_A ((\nabla \times \mathbf B) \times \mathbf {\hat {B}}) \times \mathbf {\hat {B}}\right\} \times \mathbf B\right], \\
\nabla^2 \Phi&=&4 \pi G \rho,
\end{eqnarray}
where $\rho$ is the gas density, 
$P$ is the gas pressure, 
$\magB$ is the magnetic field and
${\mathbf {\hat {B}}}\equiv \magB/|\magB| $,
$\eta_O$ and $\eta_A$ 
are, respectively, the resistivity for ohmic and ambipolar diffusion,
$E_r$ is the radiation energy, 
$\mathbf{F}_{\bm r}$ is the radiation flux,
$\mathbb{P}_r$ is the radiation pressure,
$T_g$ is the gas temperature,
$\kappa_P$ is the Plank mean opacity, 
$e=\rho u+\frac{1}{2}(\rho \mathbf{v}^2+\mathbf{B}^2)$ is 
the total energy (where $u$ is specific internal energy),
$\Phi$ is the gravitational potential,
and $a_r$ and $G$ are the radiation and
gravitational constants, respectively.
The Hall effect is not considered in this study but
is discussed in 
\citet{2011ApJ...733...54K,2011ApJ...738..180L,2015ApJ...810L..26T,
2016MNRAS.457.1037W,2017PASJ...69...95T,2018MNRAS.475.1859W}.

To close the equations for radiation transfer, 
we employed the flux-limited diffusion (FLD) 
approximation \citep{1981ApJ...248..321L},
\begin{eqnarray}
\mathbf{F}_{\bm r}&=&\frac{c\lambda}{\kappa_R \rho}\nabla E_r,\hspace{1em}
\lambda(R)=\frac{2+R}{6+2R+R^2},\nonumber\\
R&=&\frac{|\nabla E_r|}{\kappa_R \rho E_r}, \hspace{1em}
\mathbb{P}_r=\mathbb{D}E_r,\nonumber \\
\mathbb{D}&=&\frac{1-\chi}{2}\mathbb{I}+\frac{3\chi-1}{2}\mathbf{n}\otimes \mathbf{n},\hspace{1em}
\chi=\lambda+\lambda^2R^2,\hspace{1em} \nonumber \\
\mathbf{n}&=&\frac{\nabla E_r}{|\nabla E_r|},\nonumber
\end{eqnarray}
where $\kappa_R$ is the Rosseland mean opacity.

We use the smoothed particle hydrodynamics (SPH) method 
\citep{1977AJ.....82.1013L,1977MNRAS.181..375G,1985A&A...149..135M},
the numerical code was developed
by the present authors and employed previously
\citep[e.g.,][]{2011MNRAS.416..591T,2013MNRAS.428.1321T,
2013MNRAS.436.1667T,2015ApJ...810L..26T,2016ApJ...833..105Y}.
The ideal MHD part is solved with the 
Godunov smoothed particle magnetohydrodynamics (GSPMHD) method 
\citep{2011MNRAS.418.1668I}.
The divergence-free condition is maintained with
the hyperbolic divergence cleaning method for 
GSPMHD \citep{2013ASPC..474..239I}.
Radiative transfer is solved implicitly using the method of
\citet{2004MNRAS.353.1078W,2005MNRAS.364.1367W}.
We treat ohmic and ambipolar diffusions with the methods of 
\citet{2013MNRAS.434.2593T} and
\citet{2014MNRAS.444.1104W}, respectively.
Both diffusion processes were 
accelerated by super-time stepping (STS)
\citep{Alexiades96}.
To calculate the self-gravity, we adopted the Barnes-Hut tree algorithm
with an opening angle of $\theta_{\rm gravity}=0.5$ \citep{1986Natur.324..446B}. 
For the dust and gas opacities, the data tables from \citet{2003A&A...410..611S} 
and from \citet{2005ApJ...623..585F}, respectively, were adopted.
We adopted the tabulated equation of state (EOS) used in
\citet{2013ApJ...763....6T}, in which the internal degrees of freedom 
and chemical reactions of seven 
species ${\rm H_2,~H,~H^+,~He,~He^+,He^{++}, e^-}$ are considered.

We used tabulated resistivity 
calculated by the methods 
described in \citet{2002ApJ...573..199N} and \citet{2009ApJ...698.1122O}
in which we considered ion species of
${\rm H_3^+,~HCO^+,~Mg^+,~He^+,C^{+},H^+,  e^-}$.
The reaction rate was taken from the UMIST2012 database \citep{2013A&A...550A..36M}.
We took into account the cosmic ray and thermal ionization, gas-phase 
and dust-surface recombination, and ion-neutral reaction.
The dust size and density are $a=3.5 \times 10^{-2} ~{\rm \mu m}$ 
and $\rho_d=2 \gcm$, respectively. 
We assumed a dust-to-gas mass ratio of 1.7 \% to mimic
the gas of the minimum-mass solar nebula model \citep{1981PThPS..70...35H}.
Our calculations considered non-thermal ionization by the cosmic ray and the thermal 
ionization of potassium. The cosmic-ray ionization rate was fixed 
to be $\xi_{\rm CR}=10^{-17} s^{-1}$.

\subsection{Initial and boundary conditions}
We modeled the initial cloud cores as isothermal uniform gas spheres.
The mass and temperature of the initial core
are 1 $M_\odot$ and 10 K, respectively.
The rotation energy normalized by the gravitational energy of the initial core
is $E_{\rm rot}/E_{\rm grav}=0.03$.
The initial angular-momentum vector is parallel to the $z$ axis.
The initial mass-to-flux ratio 
relative to the critical value 
is $\mu=(M/\Phi)/(M/\Phi)_{\rm crit}=4$ where
$\Phi=\pi R_{\rm core}^2 B_0$ ($R_{\rm core}$ is the core radius and $B_0$ is 
the initial magnetic field strength), and 
$(M/\Phi)_{\rm crit}=(0.53/3 \pi)(5/G)^{1/2}$ \citep{1976ApJ...210..326M}.
The initial magnetic field is uniform and tilted relative to the $z$ axis
and is given by
\begin{eqnarray}
\magB=\left(B_x,B_y,B_z \right)=B_0\left( -\sin \theta,0,\cos \theta \right).
\end{eqnarray}
In the following simulations, we scanned over the parameter space comprising
two parameters: 
the ratio $\alpha$ of the thermal energy
to the gravitational energy ($\alpha \equiv E_{\rm therm}/E_{\rm grav}$)
and the relative angle $\theta$ between the magnetic 
field and angular momentum of the cloud core.
In practice, different values of $\alpha$ 
are realized by changing the core radius,
while the core mass is fixed to $1\msun$.
Thus, the rotation energy is also different between cores with different value of $\alpha$.
Table 1 lists the model names, $\alpha$, the initial radius, 
the initial density, the initial angular velocity, the initial magnetic field strength,
and  $\theta$ for the cloud core.
For comparison, we also simulated the cores with
very weak magnetic field where the mass-to-flux ratio was $\mu=10^4$ 
for $\alpha=0.2,~0.4,~0.6$, $\theta=0^\circ$
which essentially correspond to the hydrodynamics simulations.
Thus, we refer to these simulations as hydrodynamics simulations
(results are shown in figure \ref{evolution_MHD_J_sphere}).
The initial cores were modeled with $3 \times 10^6$ SPH particles.
We conducted the simulations until the epoch just after protostar formation
(when the central density $\rho_c$ becomes $\sim 10^{-2} \gcm$).

A boundary condition was imposed at $R_{\rm out}=0.95 R_{\rm core}$ and
the particles with $r>R_{\rm out}$ rotated with the initial
angular velocity. 
In addition, another boundary condition for radiative
transfer was introduced by fixing the gas and radiation temperatures
to 10 K in $\rho<4.0 \times 10^{-17} \gcm$. 

\begin{table*}
\label{initial_conditions}
\begin{center}
\caption{Model names and simulation parameters:
$\alpha=E_{\rm therm}/E_{\rm grav}$
the initial radius $R_0$, the initial density $\rho_0$, the initial angular velocity $\Omega_0$,
the initial magnetic field strength $\magB_0$,
the angle between the initial magnetic field and the 
initial angular momentum vector of cloud cores $\theta$.
The last column indicates whether the ohmic and ambipolar diffusions are
included ("Yes") or not (``No").
}		
\begin{tabular}{ccccccccc}
\hline\hline
 Model name  & $\alpha$ & $R_0 [{\rm AU}]$& 
density $\rho_0 [{\rm\gcm}]$ & $\Omega_0 [{\rm s^{-1}}]$ &$\magB_0 [{\rm G}]$ & $\theta [{\rm deg}]$  & magnetic diffusion \\
\hline
I1 & 0.2&$2.0\times10^{3}$&$1.7\times10^{-17}$&$6.5\times10^{-13}$&$3.5\times10^{-4}$& $0^{\circ}$     & No \\
I2 & 0.2&$2.0\times10^{3}$&$1.7\times10^{-17}$&$6.5\times10^{-13}$&$3.5\times10^{-4}$& $45^{\circ}$     & No \\
I3 & 0.2&$2.0\times10^{3}$&$1.7\times10^{-17}$&$6.5\times10^{-13}$&$3.5\times10^{-4}$& $90^{\circ}$     & No \\
I4 & 0.4&$4.1\times10^{3}$&$2.0\times10^{-18}$&$2.3\times10^{-13}$&$8.6\times10^{-5}$& $0^{\circ}$     & No \\
I5 & 0.4&$4.1\times10^{3}$&$2.0\times10^{-18}$&$2.3\times10^{-13}$&$8.6\times10^{-5}$& $45^{\circ}$     & No \\
I6 & 0.4&$4.1\times10^{3}$&$2.0\times10^{-18}$&$2.3\times10^{-13}$&$8.6\times10^{-5}$& $90^{\circ}$     & No \\
I7 & 0.6&$6.1\times10^{3}$&$6.1\times10^{-19}$&$1.2\times10^{-13}$&$3.8\times10^{-5}$& $0^{\circ}$     & No \\
I8 & 0.6&$6.1\times10^{3}$&$6.1\times10^{-19}$&$1.2\times10^{-13}$&$3.8\times10^{-5}$& $45^{\circ}$     & No \\
I9 & 0.6&$6.1\times10^{3}$&$6.1\times10^{-19}$&$1.2\times10^{-13}$&$3.8\times10^{-5}$& $90^{\circ}$     & No \\
\hline
NI1 & 0.2&$2.0\times10^{3}$&$1.7\times10^{-17}$&$6.5\times10^{-13}$&$3.5\times10^{-4}$& $0^{\circ}$     & Yes \\
NI2 & 0.2&$2.0\times10^{3}$&$1.7\times10^{-17}$&$6.5\times10^{-13}$&$3.5\times10^{-4}$& $45^{\circ}$     & Yes \\
NI3 & 0.2&$2.0\times10^{3}$&$1.7\times10^{-17}$&$6.5\times10^{-13}$&$3.5\times10^{-4}$& $90^{\circ}$     & Yes \\
NI4 & 0.4&$4.1\times10^{3}$&$2.0\times10^{-18}$&$2.3\times10^{-13}$&$8.6\times10^{-5}$& $0^{\circ}$     & Yes \\
NI5 & 0.4&$4.1\times10^{3}$&$2.0\times10^{-18}$&$2.3\times10^{-13}$&$8.6\times10^{-5}$& $45^{\circ}$     & Yes \\
NI6 & 0.4&$4.1\times10^{3}$&$2.0\times10^{-18}$&$2.3\times10^{-13}$&$8.6\times10^{-5}$& $90^{\circ}$     & Yes \\
NI7 & 0.6&$6.1\times10^{3}$&$6.1\times10^{-19}$&$1.2\times10^{-13}$&$3.8\times10^{-5}$& $0^{\circ}$     & Yes \\
NI8 & 0.6&$6.1\times10^{3}$&$6.1\times10^{-19}$&$1.2\times10^{-13}$&$3.8\times10^{-5}$& $45^{\circ}$     & Yes \\
NI9 & 0.6&$6.1\times10^{3}$&$6.1\times10^{-19}$&$1.2\times10^{-13}$&$3.8\times10^{-5}$& $90^{\circ}$     & Yes \\
\hline
\hline
\end{tabular}
\end{center}
\footnotesize
\end{table*}

\section{Results}
\label{results}

\subsection{Density and angular momentum structures}
\label{ret_1}

\subsubsection{Ideal MHD simulations}
\begin{figure*}
\vspace{0.5cm}
\includegraphics[width=150mm]{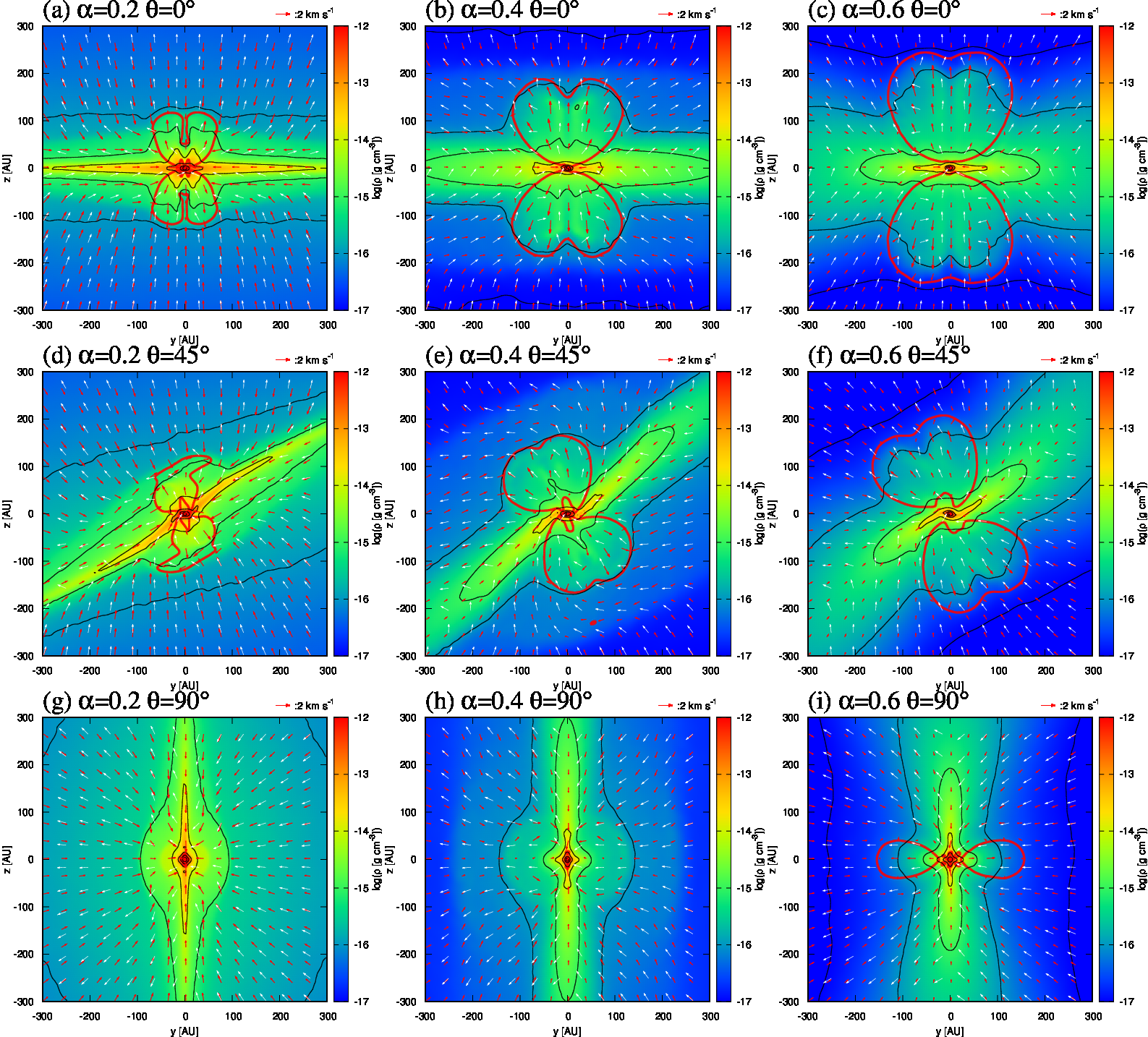}
\caption{
Density cross sections in the $y$-$z$ plane for the
central 600 AU square region
at the end epoch 
(at the epoch when the central density has reached $\rho_c \sim 10^{-2} \gcm$ ) in the simulations
of the ideal MHD models (Model I1-I9).
The black and red lines represent the contours of 
the density and of $v_r \equiv (yv_y+zv_z)/\sqrt{y^2+z^2}=0$, respectively.
Black contour levels 
are $\rho=10^{-17},10^{-16},~10^{-15},~10^{-14},10^{-13},$ and $10^{-12}\gcm$.
The region enclosed by the red contours is the outflow.
The red arrows represent the velocity field and 
the white arrows the direction of the magnetic field.}
\label{density_map_ideal}
\end{figure*}

Figure \ref{density_map_ideal} shows the density cross section in the $y$-$z$ plane, 
obtained from the ideal MHD simulations.
In the case of a non-zero $\theta$, the system loses its axial symmetry about the $z$ axis
and the structures on the $x$-$z$ and $y$-$z$ planes, for example, are different.
In particular, the outflow is not found to lie within the $x$-$z$ plane in our results
for $\theta\neq 0^\circ$.
By examining the cross section around the $z$ axis
with $0^\circ<\phi<180^\circ$, where $\phi$ is the azimuthal angle of the 
intersection of the cross section on the $x$-$y$ plane
and $\phi=0^\circ$ corresponds to the cross section of 
the $x$-$z$ plane,  we find that the outflow
sit on or close to the $y$-$z$ plane (or a plane with $\phi=90^\circ$) 
with our initial conditions.
Thus, we chose the $y$-$z$ plane to investigate 
the density and velocity structures.

Figure \ref{density_map_ideal} clearly shows two 
notable structures: the pseudo-disk and the outflow.
The pseudo-disk has the morphology of a flattened disk.
The inclination of its normal vector relative to the $z$ axis
corresponds roughly to the value of $\theta$ in the simulations.
At the beginning of the isothermal collapse phase, 
the collapse is spherically symmetric
and the central magnetic field is amplified according to
$B\propto\rho^{2/3}$.
The spherical collapse continues for as long as 
the Lorentz force is negligible.
Once the Lorentz force has become comparable to the gravity force,
it begins 
to deflect the gas motion in the direction parallel to the magnetic field and
the spherical symmetry of the isothermal collapse breaks down.
As a result, the fluid elements at a certain direction 
selectively accrete onto the midplane
and the flattened disk-like structure, i.e., pseudo-disk is formed.
Thus, the existence of a pseudo-disk or flattened envelope indicates that
non-spherical collapse has occurred in the isothermal collapse phase
\citep[see,][for detailed evolution of the
magnetic field strength]{2015MNRAS.452..278T}.
As shown below, the non-spherical collapse affects 
the evolution of the angular momentum of the central region and the early evolution of the disk.

Figure \ref{density_map_ideal} also
shows contours of $v_r \equiv (yv_y+zv_z)/\sqrt{y^2+z^2}=0$ (in red).
The enclosed regions are outflows, which
are observed to be formed in all 
the simulations with $\theta=0^\circ$ and $45^\circ$.
The outflow size decreases as $\alpha$ decreases
because the protostar quickly forms in the simulations with 
a small $\alpha$ and
there is not enough time for the outflow to grow.
The outflow is generated by the rotation of the central region
and has the angular momentum extracted from the central region
\citep{2002ApJ...575..306T,2008ApJ...676.1088M}.
Figure \ref{J_map_ideal} 
shows the spatial map of the angular momentum per unit volume 
$J_{\rm yz}=(J_{\rm y}^2+J_{\rm z}^2)^{1/2}$.
$J_{\rm yz}$ does not include the contribution of the 
rotation in the $y$-$z$ plane.
The figure shows that the outflow has greater angular momentum
than the infalling region surrounding it (as 
is particularly clear in panels (c) and (f)). 
Hence, this demonstrates that the angular momentum 
has been extracted from the central region.
The outflow size decreases as $\alpha$ decreases,
whereas the angular momentum per unit volume increases
(note that the color scale increases as $\alpha$ decreases).
Interestingly, the outflow is found to form in model I9
(panel (i) of figure \ref{density_map_ideal} and \ref{J_map_ideal}).
This is consistent with the spiral-flow-type outflow reported by \citet{2017ApJ...839...69M}.

In the cores with $\theta=45^\circ$ (panels (d), (e), and (f)),
the central structure is warped and its normal is not parallel to the 
$z$-axis (the direction of the initial angular momentum).
This indicates that magnetic braking changes the direction of the 
angular momentum. \citet{2004ApJ...616..266M} reported that the 
magnetic field changes the direction of the angular 
momentum during gravitational collapse.
The formation of warped structure is consistent with their result.

\subsubsection{Non-Ideal MHD simulations}
Figures \ref{density_map} and \ref{J_map} 
show the density and angular-momentum cross section in the $y$-$z$ plane,
respectively, as obtained with the non-ideal MHD simulations.
The pseudo-disks are also formed in the non-ideal MHD simulations,
indicating that the radial magnetic tension causes a deviation of the gas motion 
from the spherical symmetric collapse also in the non-ideal MHD simulations.
This is well explained by the fact that 
magnetic diffusions do not play a role at low densities
$\rho \lesssim 10^{-13}\gcm$ \citep{1990MNRAS.243..103U,
2002ApJ...573..199N,2015ApJ...801..117T, 2015MNRAS.452..278T}.

An outflow appears in the simulations with $\theta=0^\circ$ and $45^\circ$.
The inclusion of the magnetic diffusions in the simulation results
in the weaker outflow activity.
The outflow velocity and angular momentum per unit volume
(figure \ref{J_map}) 
are smaller than those in the ideal MHD simulations.
Their size is, however, larger than in the ideal MHD simulations.
This reflects the fact that, in the non-ideal MHD simulations,
the time span after from the first-core formation
up to the second collapse is longer than in the ideal MHD simulations,
owing to the rotation support.
Thus, the system has a longer amount of time to develop a
larger outflow (see, figure \ref{time_rhoc}).

\begin{figure*}
\vspace{0.5cm}
\includegraphics[width=150mm]{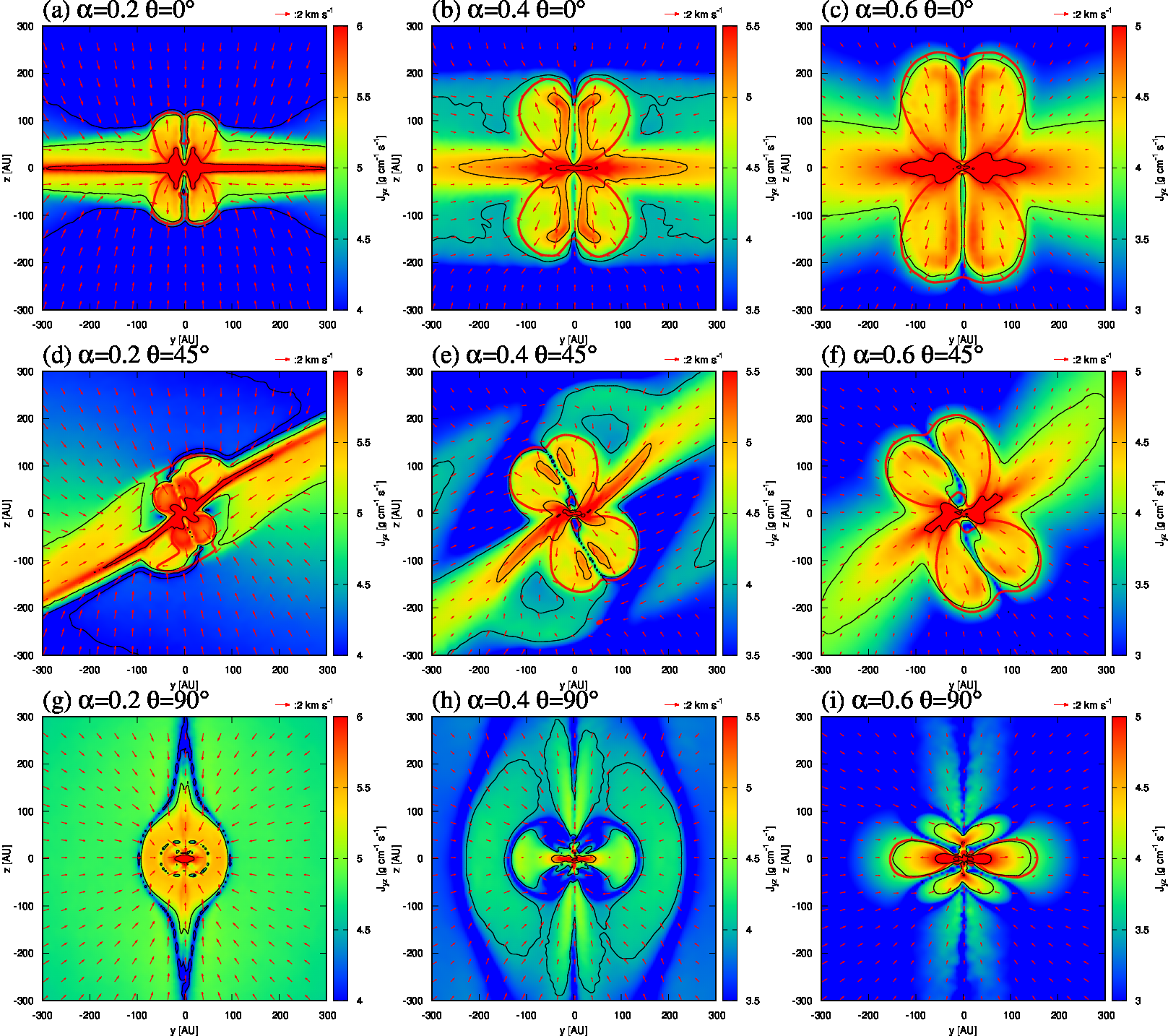}
\caption{Cross sections of the angular momenta per unit volume 
$J_{\rm yz}=(J_{\rm y}^2+J_{\rm z}^2)^{1/2}$  in the $y$-$z$ plane
for the central 600 AU square region at the end epoch 
(at the epoch when the central density has reached $\rho_c \sim 10^{-2}\gcm$ )
in the simulations of the ideal MHD models (Model I1-I9).
The black and red lines represent contours of $J_{\rm yz}$ of 
$v_r \equiv (yv_y+zv_z)/\sqrt{y^2+z^2}=0$, respectively.
The region enclosed by the red contours is outflow.
The black contour levels 
are $J_{\rm yz}=10^{4},10^{5}$, and $10^{6} {\rm g~cm^{-1}~s^{-1}}$.
The red arrows show the velocity field.
}
\label{J_map_ideal}
\end{figure*}

\begin{figure*}
\vspace{0.5cm}
\includegraphics[width=150mm]{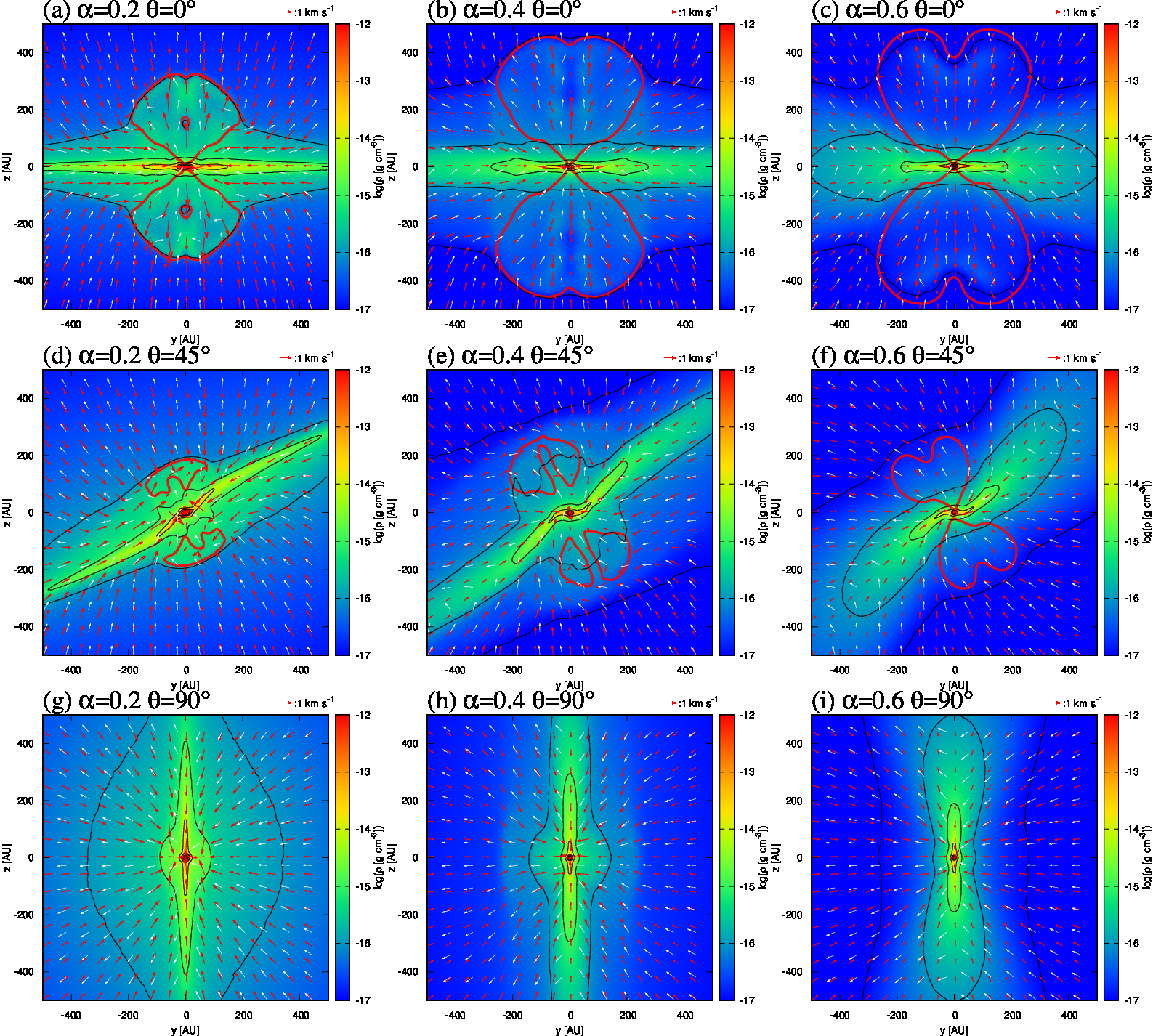}
\caption{
Density cross sections in the $y$-$z$ plane for the
central 1000-AU square region
at the end epoch 
(at the epoch when the central density has reached $\rho_c \sim 10^{-2}\gcm$ ) in the simulations of the non-ideal MHD models (Model NI1-NI9).
The black and red lines represent contours of the density and of $v_r \equiv (yv_y+zv_z)/\sqrt{y^2+z^2}=0$, respectively.
The black contour levels 
are $\rho=10^{-17},10^{-16},~10^{-15},~10^{-14},10^{-13},$ and $10^{-12}\gcm$.
The region enclosed by the red contours is outflow.
The red arrows represent the velocity field and the
white arrows the direction of the magnetic field.
}
\label{density_map}
\end{figure*}

\begin{figure*}
\vspace{0.5cm}
\includegraphics[width=150mm]{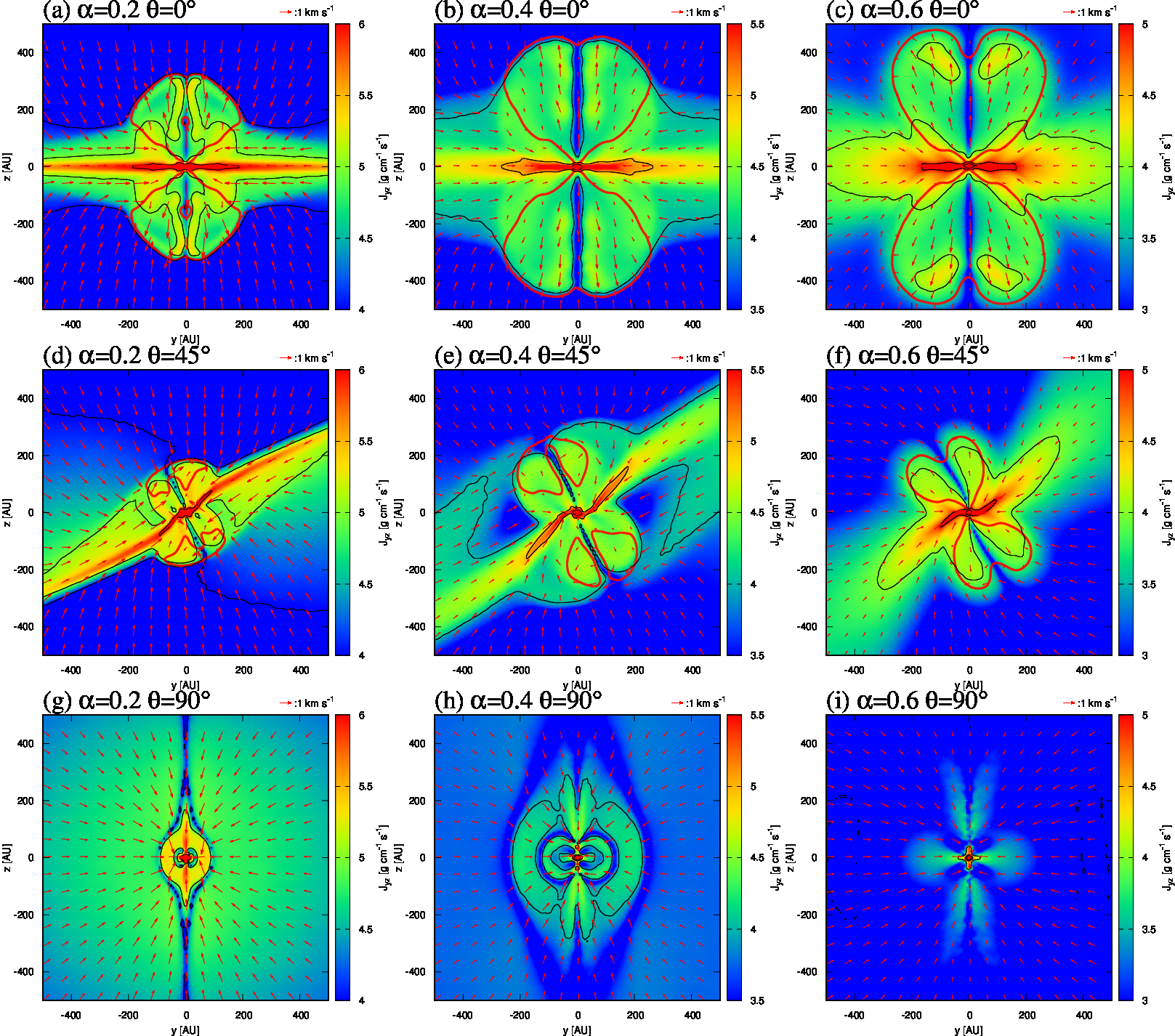}
\caption{
Cross section of the angular momenta 
per unit volume $J_{\rm yz}=(J_{\rm y}^2+J_{\rm z}^2)^{1/2}$ in
the $y$-$z$ plane for the central 1000-AU 
square region at the end epoch 
(at the epoch when the central density has reached $\rho_c \sim 10^{-2}\gcm$ ) of 
the non-ideal MHD simulations  (Model NI1-NI9).
The black and red lines represent contours of the  density and $v_r \equiv (yv_y+zv_z)/\sqrt{y^2+z^2}=0$, respectively.
The region enclosed by the red contours is the outflow.
The black contour levels 
are $J_{\rm yz}=10^{4},10^{5}$, and $10^{6} {\rm g~cm^{-1}~s^{-1}}$.
The red arrows represent the velocity field.
}
\label{J_map}
\end{figure*}

Figures \ref{J_map_ideal} and figure \ref{J_map} indicate that 
the morphology of the angular-momentum map is similar for the 
ideal and non-ideal MHD simulations although strength differ
On the other hand, the outflow morphology is different.
For example, the panels (d) and (e) of figures \ref{J_map_ideal} 
and figure \ref{J_map} show that
the gas in the central region of $|y|,|z|<100 {\rm AU}$ 
in non-ideal MHD simulations 
is not outflowing, whereas that of the ideal MHD simulation is outflowing.
Panels (i) of figures \ref{J_map_ideal} and \ref{J_map} show that
a spiral-flow type outflow does not appear in the non-ideal MHD simulations.
These results show that the outflow activity depends 
on the strength of the rotation in the envelope, and on whether or not
magnetic diffusion is considered
\citep[see also][for the relation between the outflow activity and magnetic diffusions]
{2015MNRAS.452..278T,2016A&A...587A..32M}.


\subsection{Evolution of the angular momentum of the central region}
\label{ret_2}
\subsubsection{Ideal MHD simulations}
\begin{figure*}
\includegraphics[width=40mm,angle=-90]{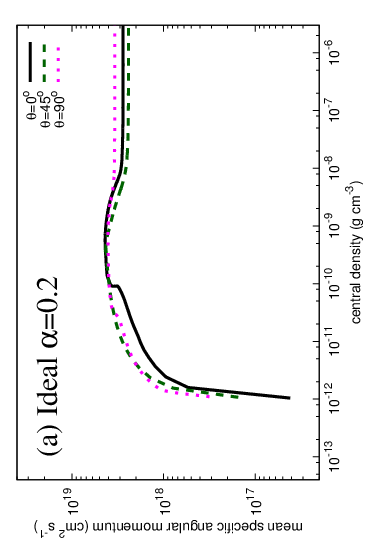}
\includegraphics[width=40mm,angle=-90]{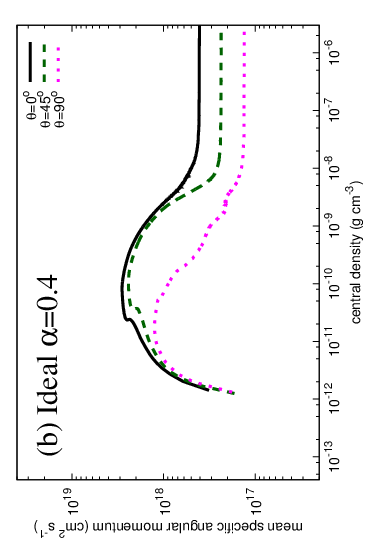}
\includegraphics[width=40mm,angle=-90]{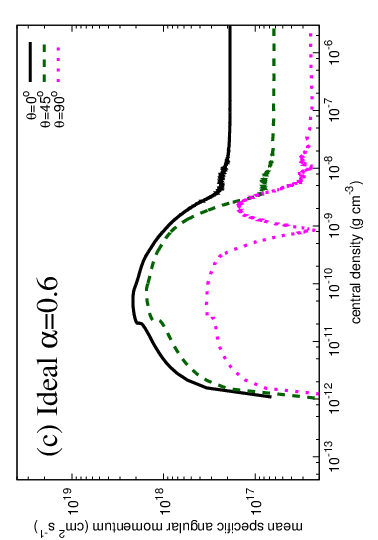}
\caption{
The evolution of 
the mean specific angular momentum in the region with
$\rho>10^{-12} \gcm$ as a function of the central density
in the ideal MHD simulations (Model I1-I9).
The horizontal axis represents the central density $\rho_c$ 
which is an increasing function of time (see figure \ref{time_rhoc}).
The vertical axis represents the mean specific 
angular momentum of the region with $\rho>10^{-12} \gcm$, where 
the mean specific angular momentum is calculated as 
$\bar{j}(\rho)=\int_{\rho>10^{-12}\gcm} \rho \mathbf{r\times v}d\mathbf{V}/\int_{\rho>10^{-12} \gcm} \rho d\mathbf{V}$. 
}
\label{rhoc_JM12_ideal}
\end{figure*}

Figure \ref{rhoc_JM12_ideal}
shows the evolution of the mean angular momentum of 
the central region (the region with $\rho>10^{-12} \gcm$) of ideal MHD simulations.
The horizontal axis represents the central density $\rho_c$.
Because $\rho_c$ is an increasing function of time 
(see, figure \ref{time_rhoc}), 
figure \ref{rhoc_JM12_ideal} shows the time evolution of the
mean angular momentum around the center.

In the models with $\alpha=0.6$ (panel (c)), which 
is a similar value to that adopted in \citet{2004ApJ...616..266M}, 
the angular momentum at the end of the 
simulation (at the epoch when the central 
density has reached $\rho_c \sim 10^{-2}\gcm$ ) is a decreasing function of the relative angle $\theta$.
This suggests that
magnetic braking is stronger in a cloud core with $\theta=90^\circ$
than $\theta=0^\circ$. 
This result is consistent with 
\citet{2004ApJ...616..266M} and with the
analytic argument of \citet{1985A&A...142...41M}.

As $\alpha$ decreases,
the difference in the angular momenta becomes smaller,
and in the models with $\alpha=0.2$, 
the central angular momentum of the core with  $\theta=90^\circ$
is slightly greater
than $\theta=0^\circ$ (right panel).
This is consistent with the results of
\citet{2009A&A...506L..29H} and \citet{2012A&A...543A.128J},
in which the initial cloud core had $\alpha=0.25$.
We note that, as in \citet{2012A&A...543A.128J},
the difference of the central angular momentum is small at the epoch
of the protostar formation.
This suggests that the evolution of the central angular momentum in the 
misaligned cloud core depends on
$\alpha$ when the ideal MHD approximation is adopted.

The figure shows that the mean angular momentum increases
in the density range of $10^{-12}\gcm \lesssim \rho_c  \lesssim 10^{-10}\gcm$.
This increase is caused by the mass and angular momentum 
accretion from the envelope.
Then, it begins to decrease 
in the range of $10^{-10}\gcm \lesssim \rho_c \lesssim 10^{-8}\gcm$.
The decrease of the central angular momentum is caused by the removal of angular momentum 
from the central region as a result of magnetic braking. This
is significant in the simulations with a large $\alpha$.
because the mass and angular momentum accretion rates decrease
as $\alpha$ increases, and because the removal of angular momentum 
by the magnetic field overtakes the 
angular momentum supply in the simulations with a greater $\alpha$ value.

By comparing the simulations with the same $\theta$ values
(i.e., lines of the same colors in the panels of
figure \ref{rhoc_JM12_ideal}),
we observe that the central angular 
momentum is a decreasing function of $\alpha$.
This result suggests that the circumstellar disk forms more easily in 
the core when $\alpha$ is small,
consistent with previous studies
\citep{2014MNRAS.438.2278M,2016MNRAS.463.4246M}.



\subsubsection{Non-Ideal MHD simulations}
Figure \ref{rhoc_JM12_nonideal}
shows the evolution of the mean angular momentum of the central region 
with $\rho>10^{-12} \gcm$ in the non-ideal MHD simulations.
In all the models, the central angular momentum decreases as $\theta$ increases.
In particular, comparing the results between the cases of
$\theta=0^\circ$ and $\theta=90^\circ$, 
we find that the central angular momentum with $\theta=0^\circ$
is greater than with $\theta=90^\circ$, independently of
$\alpha$.

Because the magnetic diffusions are effective in the relatively high
density of $\rho \gtrsim 10^{-13} \gcm$,
the suppression of magnetic braking in the high-density 
region causes the difference between the ideal and non-ideal MHD simulations.
The decrease of the angular momentum for
$\rho_c>10^{-10}\gcm$ as observed in the ideal MHD simulations (figure \ref{rhoc_JM12_ideal}),
does not appear once magnetic diffusions have been considered.
This result indicates that the removal of angular momentum from the central region
is not significant, and that supply of angular momentum by mass accretion 
dominates its removal.

Figure \ref{rhoc_JM12_nonideal} also shows that the differences in
the central angular momentum between the models with 
$\theta=0^\circ$ and $\theta=45^\circ$  are small.
For all values of $\alpha$, 
the central angular momenta differ between 
the models with $\theta=0^\circ$ and $\theta=45^\circ$
are by factor of two or less.
This is also true in the ideal MHD simulations (see, figure \ref{rhoc_JM12_ideal}).
Thus, we conclude that a misalignment with a small $\theta$ 
($\theta<45^\circ$) does not significantly affect
the evolution of the central angular momentum regardless of whether non-ideal effects 
are included or not.
The small difference in angular momentum with $\theta<45^\circ$
is consistent with a previous study \citep{2016A&A...587A..32M},
where the authors
investigated the evolution of the angular momentum in 
the core with $\theta<40^\circ$ 
and reported that the misalignment does not 
introduce any significant difference in the angular momentum
in the disk if magnetic diffusions are taken into account.

In the non-ideal MHD simulations, 
the central angular momentum is also a decreasing function of 
$\alpha$ for a given  $\theta$,
and consistent with the results of ideal MHD simulations.
This suggests that the difference in  $\alpha$ mainly affects 
magnetic braking in the isothermal-collapse phase ($\rho_c<10^{-13} \gcm$)
in which the magnetic diffusion does not play a role.

\subsubsection{Evolution of the central density}
In figure \ref{rhoc_JM12_ideal} and \ref{rhoc_JM12_nonideal},
The central density was chosen as a variable describing temporal evolution.
A similar parameterization has been commonly 
used \citep[e.g.,][]{2004ApJ...616..266M,
2014MNRAS.437...77B,2015ApJ...801..117T}.
This allows us to compare simulations at 
similar evolutionary stage, for example, at the epochs of the first- and 
($\rho_c \sim 10^{-13}\gcm$) and second-core (or protostar) formation 
($\rho_c \sim 10^{-2}\gcm$).
However, the elapsed time of 
each evolutionary epoch differs between the simulations.
We here investigate the difference in the
temporal evolution of the central density.

Figure \ref{time_rhoc} shows the temporal evolution of the central density.
The time at the second collapse is greater than the free-fall time by a 
factor between $1.1$ and $1.4$.
where the free-fall time is $t_{\rm ff}=1.6\times10^4$ yr for
$\alpha=0.2$, $t_{\rm ff}=4.6\times10^4$ yr for $\alpha=0.4$, and 
$t_{\rm ff}=8.5\times10^4$ yr for $\alpha=0.6$.
The figure shows that the epochs of the first-
($\rho_c\sim 10^{-13} \gcm$) 
and second-core formations ($\rho_c\sim 10^{-2} \gcm$) 
depend on $\theta$, even for the same $\alpha$.
In the ideal MHD simulations, the models with a greater
$\theta$ require a longer time for the 
first- and second-core formations.
For example, in the ideal MHD simulations with $\alpha=0.6$, 
the second-core formation epoch
differs by $\sim5\times10^3$ yr between the simulations
with $\theta=0^\circ$ and $\theta=90^\circ$.
This introduces complexity when we compare the simulations 
that have different $\theta$ value even when their initial cores had
the same $\alpha$ and $\beta_{\rm rot}$.

Figure \ref{time_rhoc} also shows the evolution of the central density
in the non-ideal MHD simulations.
For all the values of  $\alpha$, 
the epoch of the first-core formation ($\rho_c\sim10^{-13}\gcm$) is delayed as 
$\theta$ increases, consistent with the results of the ideal MHD simulations.
Note that the temporal evolution of $\rho_c$ is almost identical in the isothermal collapse phase 
$\rho_c\lesssim 10^{-14}\gcm$ between the ideal and non-ideal MHD simulations.

The epoch of the second-core formation in the non-ideal MHD simulations 
increases as $\theta$ increases, for $\alpha=0.4$ and $0.6$,  
although the difference between
the result with $\theta=0^\circ$ and $\theta=90^\circ$ becomes smaller than that
of the ideal MHD simulations.
In the simulation with $\alpha=0.2$, it is
longer for $\theta=0^\circ$ than for $\theta=90^\circ$.
The reason for the small difference of the second collapse epoch 
when  $\alpha=0.4$ and $0.6$, and for why the second collapse 
happens earlier in the simulation with  $\theta=90^\circ$ with $\alpha=0.2$
is due to the rotational support.
Figure \ref{rhoc_JM12_nonideal} shows that 
the central region for the models with $\theta=0^\circ$ and $45^\circ$ 
has a much greater angular momentum than with $\theta=90^\circ$.
The centrifugal radius $r_{\rm cent}$ is estimated to be
$r_{\rm cent} \sim j^2/(GM)\sim 4.5 
(j/(3\times10^{19} \cmcms))^2(M/(0.1 \msun)) ~{\rm AU}$.
The centrifugal force with a specific angular momentum of the order of
$j \sim 10^{19}\cmcms$ provides 
the rotational support against self-gravity at a
radius of several AU, which corresponds to the 
radius of the first-core.
Figures \ref{rhoc_JM12_nonideal} and \ref{time_rhoc} show that the
duration of the first-core epoch ($10^{-13}\gcm <\rho_c<10^{-8}\gcm$)
is considerably longer in the models with $j \gtrsim 10^{19}\cmcms$ than
those with $j\lesssim10^{19}\cmcms$.
The difference in the second-collapse epoch arises in consequence.

It is worth noting that the order reversal of the second-collapse epoch 
observed in our simulations
has a different physical origin from that observed in 
\citet{2018arXiv180108193V}, in which the second collapse in the ideal MHD simulation 
happened later than in the non-ideal MHD simulation.
As those authors discussed, the order reversal they observed was
caused by interchange instability.

\subsubsection{Summary of section \ref{ret_2}}
\begin{figure*}
\includegraphics[width=40mm,angle=-90]{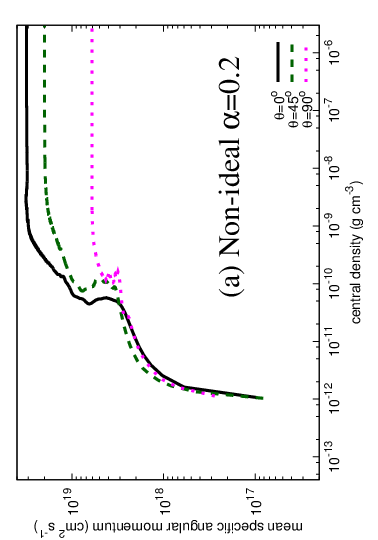}
\includegraphics[width=40mm,angle=-90]{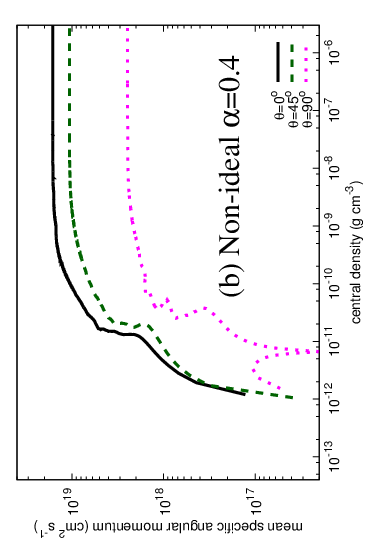}
\includegraphics[width=40mm,angle=-90]{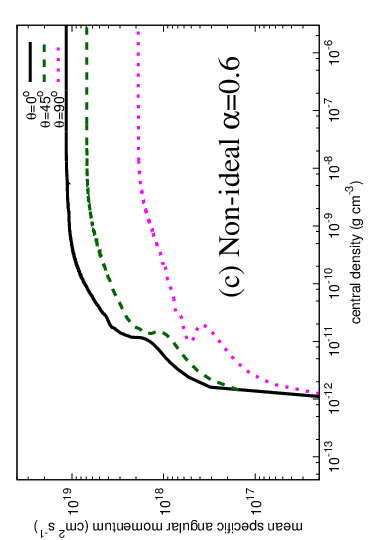}
\caption{
Evolution of 
the mean specific angular momentum in the region with
$\rho>10^{-12} \gcm$ as a function of the central density
in the non-ideal MHD simulations (Model NI1-NI9).
}
\label{rhoc_JM12_nonideal}
\end{figure*}

\begin{figure*}
\includegraphics[width=40mm,angle=-90]{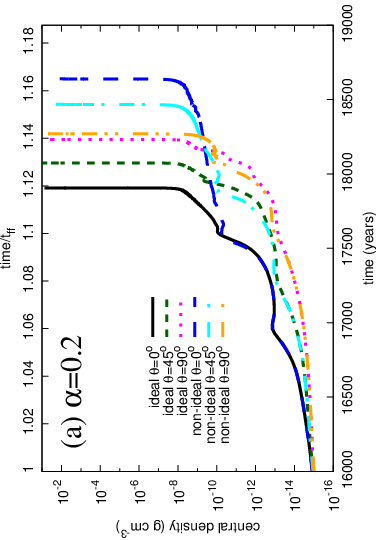}
\includegraphics[width=40mm,angle=-90]{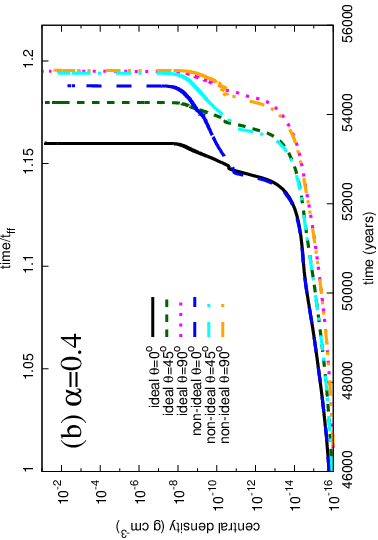}
\includegraphics[width=40mm,angle=-90]{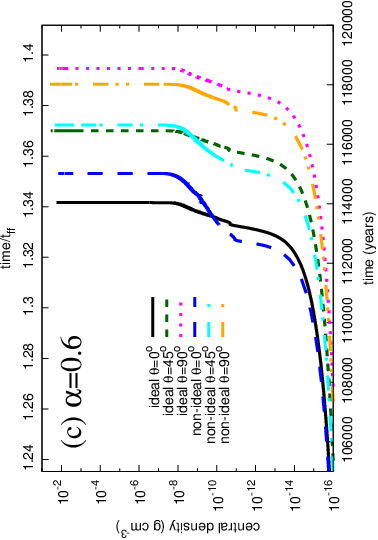}
\caption{
Evolution of the central density as a function of the elapsed time.
The scaling factor $t_{\rm ff}$ in the upper horizontal axis denotes the free-fall time.
}
\label{time_rhoc}
\end{figure*}

This subsection demonstrated that the evolution in angular momentum of the central
region depends on  the $\alpha$ value of the initial cloud cores, and
on whether magnetic diffusions are taken into account.
In particular, it was seen that 
even the qualitative dependence of the central angular momentum
on $\theta$ varies, depending on  $\alpha$ in ideal MHD simulations.
For small value of $\alpha$ such as $\alpha=0.2$,
the central angular momentum of the simulation with $\theta=0$
is smaller than that with $\theta=90^\circ$ 
(figure \ref{rhoc_JM12_ideal}).
In contrast,
For larger value of $\alpha$,
the central angular momentum of the simulation with $\theta=0$
is greater than that with $\theta=90^\circ$ 
(figure \ref{rhoc_JM12_nonideal}).
These results are consistent with previous studies 
which employed a large value of  $\alpha$  \citep{2004ApJ...616..266M} and 
a small  $\alpha$ 
\citep{2009A&A...506L..29H,2012A&A...543A.128J}.
the results becomes simpler once magnetic diffusion has been incorporated into the simulation.
The central angular momentum decreases as $\theta$ increases, regardless
of the value of $\alpha$ (figure \ref{rhoc_JM12_nonideal}).


\subsection{Angular momentum evolution of the fluid element in the central region}
\label{ret_3}
This subsection investigates the underlying mechanism that
causes the diversity of the evolution of the angular momentum 
described above.
In the previous subsection and in most earlier studies,
the temporal evolution of the central angular 
momentum of a collapsing cloud core 
was investigated by specifying threshold densities 
for the ``central region" 
\citep[e.g.,][]{2012A&A...543A.128J,2015MNRAS.452..278T}
or a criterion which determines the ``disk" \citep[][]{2016A&A...587A..32M}.
However, these kinds of analyses make
it is difficult to identify the mechanism causing the diversity
of the evolution of the angular momentum,
because the resultant angular momentum of the central region is
determined by the supply of angular momentum
provided by the envelope accretion (the angular momentum of the accretion flow
may have changed from its initial value already)
and by the removal of the angular momentum from 
the central region by the magnetic field.
Thus, it is unclear when and how the angular momentum of the
gas changes.
We therefore investigate the evolution of the angular momentum
of each fluid element during cloud-core collapse.

The following three subsections
show that three different mechanisms 
affect the evolution of the angular momentum evolution.
The first is anisotropic accretion caused by radial magnetic tension.
The second is magnetic braking during the isothermal collapse phase.
The third is magnetic braking in
the first-core or new-born disk, which plays a significant role in
the ideal MHD simulations.

\subsubsection{Initial locations of the gas falling into the central region}
\begin{figure*}
\vspace{1.5cm}
\includegraphics[width=45mm,angle=-90]{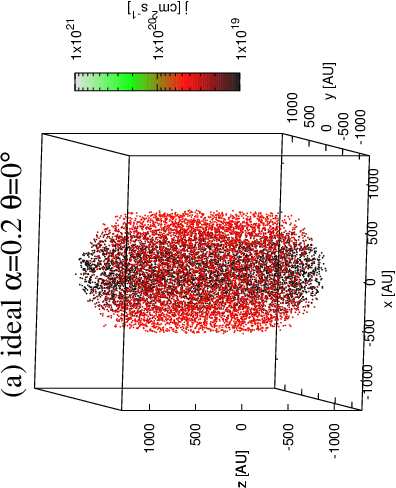}
\includegraphics[width=45mm,angle=-90]{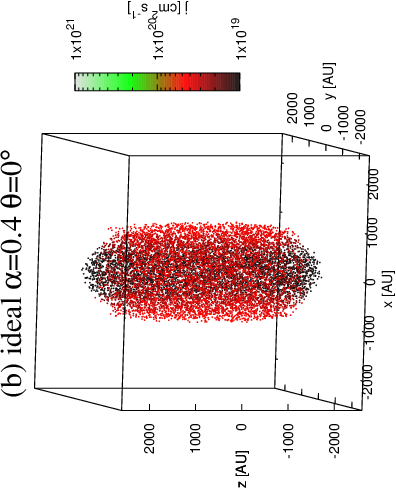}
\includegraphics[width=45mm,angle=-90]{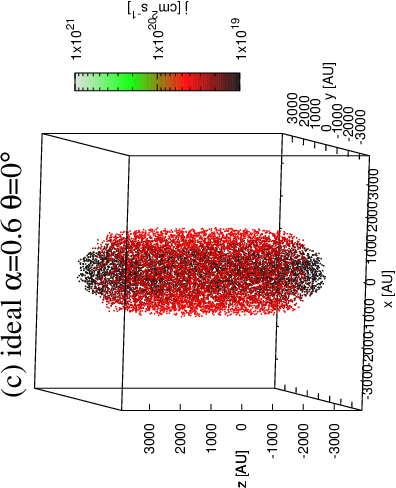}
\includegraphics[width=45mm,angle=-90]{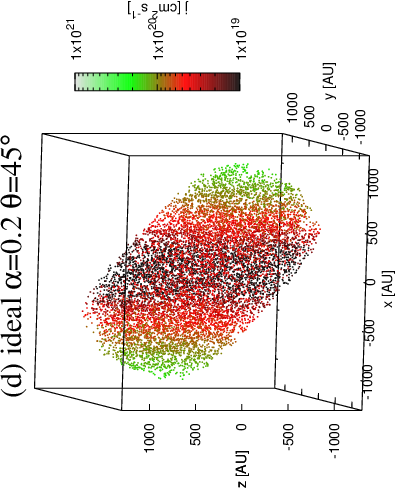}
\includegraphics[width=45mm,angle=-90]{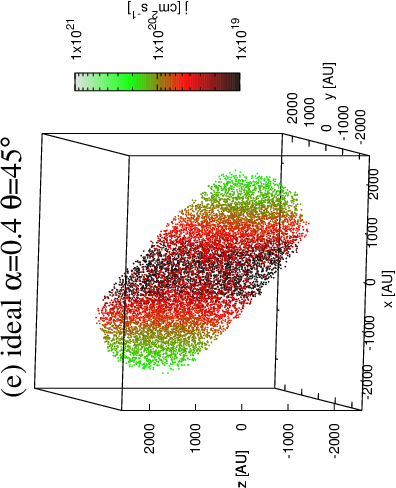}
\includegraphics[width=45mm,angle=-90]{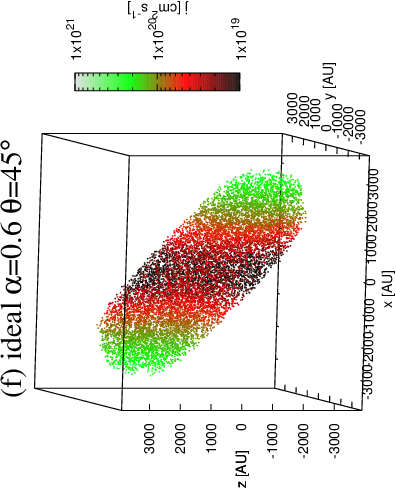}
\includegraphics[width=45mm,angle=-90]{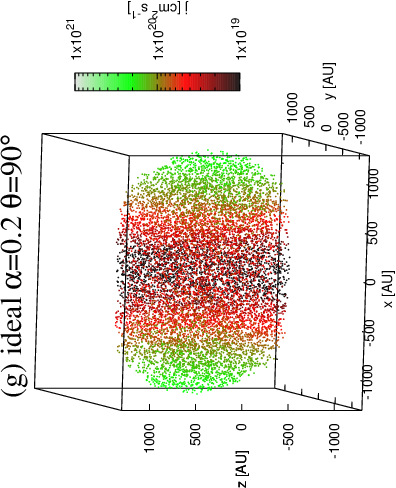}
\includegraphics[width=45mm,angle=-90]{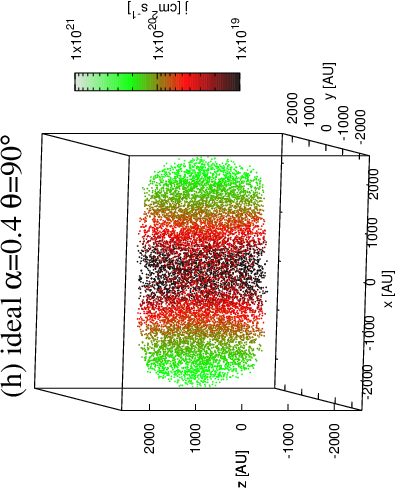}
\includegraphics[width=45mm,angle=-90]{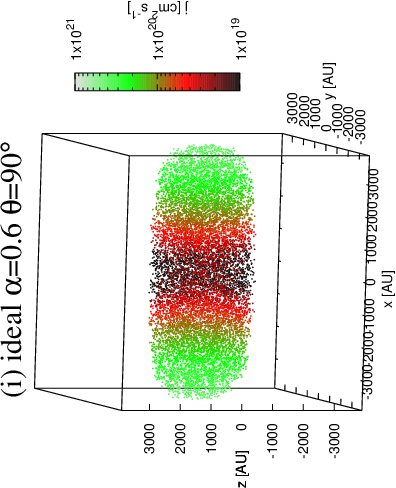}
\caption{
Initial distribution of the SPH particles that have fallen into the central region 
(with density greater than $10^{-12} \gcm$ at the end of the simulation),
obtained from the ideal MHD simulations (Model I1-I9).
The color scale shows the specific angular momentum of each particle in the initial state.
The initial angular momentum of the core is parallel to the $z$ axis, and 
the amount of specific angular momentum of 
each particle is proportional to $(x^2+y^2)$.
}
\label{init_dist_particle}
\end{figure*}

Figure \ref{init_dist_particle} shows the initial distribution of the 
fluid elements that have fallen into 
the central region by the end of the ideal MHD simulations 
(by the epoch when the central density has reached $\rho_c \sim 10^{-2}\gcm$ ). 
Following the criterion used in the previous section, 
we define the central region as the region where
$\rho>10^{-12}\gcm$. At the end of each simulation run, 
$10^4$ SPH particles were sampled  from the central region.
We then traced back their trajectories backward and determined their initial positions.

The figure shows that the distributions 
of the initial fluid-element locations
are not spherically symmetric but
prolate, with the major axis running nearly parallel to the initial magnetic field.
During cloud-core collapse, the gas tends to move along the field
line by the Lorenz force. 
thereby giving the distributions of the prolate shape.

The color of each fluid element represents the initial angular momentum.
Because we assumed that the rotation axis of the initial cloud core 
was parallel to the $z$ axis,
the angular momentum of the fluid elements is proportional to $(x^2+y^2)$,
i.e., the square of the distance from the $z$-axis.
Because the major-axis of the prolate distribution 
is found to be parallel to the initial magnetic field,
the fluid elements that initially had a small angular momentum selectively 
accreted onto the central region when  $\theta=0^\circ$ whereas
the fluid elements which initially have a large angular momentum selectively 
accreted onto the central region when $\theta=90^\circ$.
If we only investigate the central angular momentum, 
this non-spherical symmetric accretion apparently causes strong and weak magnetic 
braking in the cases with, respectively,  $\theta=0^\circ$ and $90^\circ$.
Note that this effect is ``apparent" in the sense that
the fluid elements with a large (for $\theta=0^{\circ}$) or small 
(for $\theta=90^{\circ}$) angular momentum 
eventually accrete to the central region in the subsequent evolution,
though their timing is delayed.

By comparing the panels with the same $\theta$ values (e.g., (a), (b), and (c)), 
we find that the ratio of the major-axis 
of the particle distribution decreases as 
$\alpha$ decreases, and that the 
initial distribution becomes more spherical.
A core with a small $\alpha$ corresponds to a more unstable 
cloud core. A stronger gravitational force is exerted on each fluid element,
and the Lorentz force becomes relatively weak.
As a result, the initial distribution 
is more spherical in simulations with a small $\alpha$.

\subsubsection{Evolution of the angular momentum of the fluid elements in the central region}
\begin{figure*}
\includegraphics[width=40mm,angle=-90]{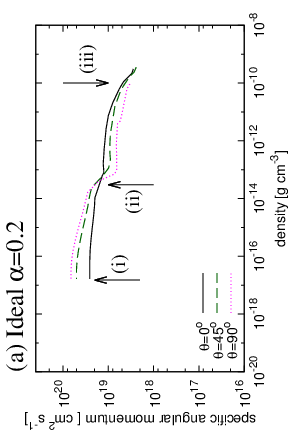}
\includegraphics[width=40mm,angle=-90]{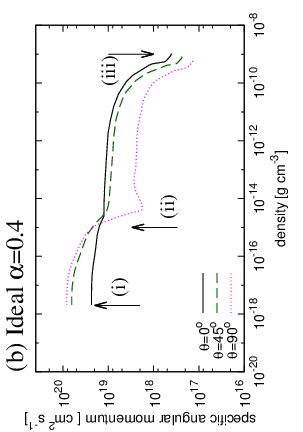}
\includegraphics[width=40mm,angle=-90]{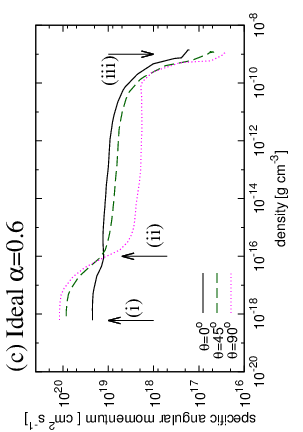}
\caption{
Temporal evolution of the mean specific angular momentum of the fluid elements 
that have fallen into the central region by the end of the ideal MHD simulations
(the particles shown in figure \ref{init_dist_particle}).
The horizontal axis represents the median density of the fluid elements,
which expresses the temporal evolution.
The vertical axis shows the mean specific angular 
momentum of the fluid elements.
The arrows indicate the positions where (i) the initial angular momentum,
(ii) magnetic braking in the isothermal collapse phase, and 
(iii) magnetic braking in the first-core and disk are found.
}
\label{evolution_ideal_MHD_J_center}
\end{figure*}

\begin{figure*}
\includegraphics[width=40mm,angle=-90]{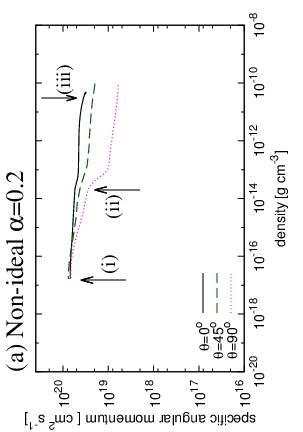}
\includegraphics[width=40mm,angle=-90]{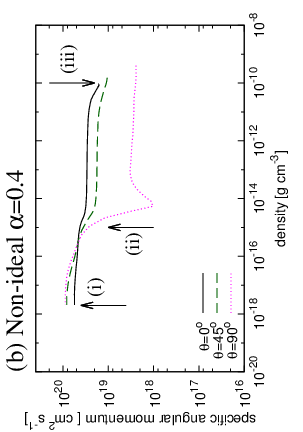}
\includegraphics[width=40mm,angle=-90]{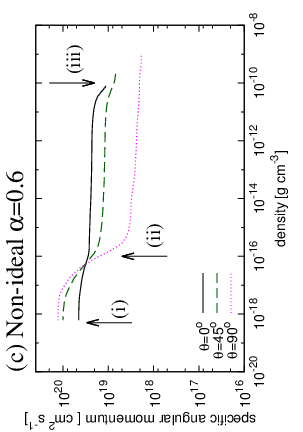}
\caption{
Temporal evolution of the mean specific angular momentum of the fluid elements 
that have fallen into the central region 
(with density greater than $10^{-12} \gcm$ at the end of the non-ideal MHD simulations).
The horizontal axis represents the median density of the fluid elements
which describes the temporal evolution.
The vertical axis shows the mean specific angular momentum of the fluid elements.
The arrows indicate the positions where (i) the initial angular momentum,
(ii)  magnetic braking in the isothermal collapse phase, and 
(iii) magnetic braking in the first-core and disk are found.
}
\label{evolution_resistive_MHD_J_center}
\end{figure*}

Figure \ref{evolution_ideal_MHD_J_center}
shows the temporal evolution of the mean specific angular momenta of
the fluid elements 
that have fallen into the central region at the end of the ideal MHD
simulations (the particles shown in figure \ref{init_dist_particle}).
The horizontal axis represents the median density 
of the fluid elements which describes the time evolution.
We use the  median density of the sampled particles 
for the following reasons.
Using figure \ref{evolution_ideal_MHD_J_center}
and \ref{evolution_resistive_MHD_J_center}, 
we want to determine the typical density where the most sampled
particles lose their angular momentum.
However, we found that the central density does not correctly 
represent the typical density.
We tried to find a good quantity for our purpose and found the 
median density to be suitable for indicating the density where 
most particles lose their angular momentum.
The reason for the initial increase in angular momentum as $\theta$ 
increases (where indicated with arrows (i) in figure \ref{init_dist_particle}))
is the non-spherical accretion discussed in the previous subsection.

Figure \ref{evolution_ideal_MHD_J_center} 
indicates that the mean specific angular momentum 
decreases in the two different density ranges.
A decrease occurs in $\rho \lesssim10^{-13}\gcm$, 
i.e., in the isothermal collapse phase (indicated with arrows (ii)).
At this stage, the angular 
momentum decreases more rapidly in the models with a larger $\theta$.
This suggests that the magnetic braking in the isothermal phase is
strong for a greater $\theta$. We will come back to this point 
in the next subsection.

The other decrease occurs in $\rho\gtrsim 10^{-11} \gcm$,
i.e., in the first-core or disk around it (indicated with arrows (iii)).
In this phase, the gas is supported by the pressure 
or rotation \citep[when the disk around the first-core is 
formed, see ][]{2015ApJ...801..117T,2015MNRAS.452..278T}.
The two-step decrease of the angular momentum
was observed in the pioneering two-dimensional ideal MHD simulation
by \citet{2000ApJ...528L..41T}.
He showed that approximately  70\% of the angular momentum is extracted during the
isothermal-collapse phase and that the remaining angular momentum is extracted 
during the adiabatic-collapse phase.
The results of our ideal MHD simulations with $\theta=0^\circ$ are
consistent with the results obtained in \citet{2000ApJ...528L..41T}.

Figure \ref{evolution_ideal_MHD_J_center} also shows that 
the decreases in angular momentum in both the isothermal and adiabatic phases
are less significant in a cloud core with a small $\alpha$.
Because gravity is strong in the core 
with a small $\alpha$ and
free-fall time is short, there is not enough time 
for the magnetic field to extract the angular momentum 
from the fluid elements.
The results again suggest that $\alpha$  
(or the degree of gravitational instability of the initial core) is
an important parameter for investigating the effect of magnetic braking 
in collapsing cloud cores.


Figure \ref{evolution_resistive_MHD_J_center} shows the evolution of the mean specific 
angular momentum for the 
central fluid elements in the non-ideal MHD simulations.
As in the case of figure \ref{evolution_ideal_MHD_J_center},
$10^4$ SPH particles were sampled from the region with
$\rho>10^{-12}\gcm$ at the end of the simulation and 
Their trajectories were traced back to determine their initial positions.
Note that the sampled particles are not the same in
the ideal and non-ideal MHD simulations for given values of $\alpha$
and $\theta$, despite starting from the same 
initial cloud core. This is  because the structures of the central region 
at the end of the simulations are different as a result of inclusion of  the magnetic diffusion.

Figure \ref{evolution_resistive_MHD_J_center} shows that the decrease in angular momentum 
in the isothermal-collapse 
phase ($\rho\lesssim10^{-13} \gcm$)
also occurs in non-ideal MHD simulations 
(indicated with arrows (ii)) and that the mean angular momentum 
in the $\theta=90^\circ$ model becomes
smaller than that for $\theta=0^\circ$  in the isothermal phase in all simulations.
The evolution of the angular momentum in the isothermal phase 
is similar to that obtained with the ideal MHD approximation because 
magnetic diffusion is effective only for $\rho \gtrsim 10^{-13} \gcm$.

Unlike the results of the ideal MHD simulations, the secondary decrease in 
the angular momentum (indicated with arrows (iii)) is weak.
The ohmic and ambipolar diffusion efficiently removes the magnetic flux from the
region with $\rho\gtrsim10^{-13} \gcm$ and the magnetic field is weaker
in the central region \citep[see, figures 3 and 4 of][]{2015MNRAS.452..278T} than
in the ideal MHD simulations. 
Furthermore, the gas and the magnetic field are almost 
decoupled in the central region and the toroidal magnetic 
field generated by the gas rotation, which causes 
the magnetic torque is also weak.
As a result, the efficiency of magnetic braking
is reduced in the high-density region compared to the ideal MHD simulations.




\subsubsection{Evolution of the angular momentum of 
the fluid elements initially in the spherical region}
The initial distributions of the fluid elements discussed in
the previous section differ between models because the
fluid elements are sampled at the end of the simulations
(figure \ref{init_dist_particle}).
Thus, although we have shown that the angular momentum significantly
decreases in the isothermal-collapse phase in the simulations 
with $\theta=90^\circ$, it remains debatable whether 
the magnetic braking in the isothermal-collapse phase is stronger
in the perpendicular than in the parallel configuration.
This is because the initial distribution discussed in the previous section
is elongated along the initial magnetic field 
and has a longer lever arm in the core with a greater $\theta$.
The long and small lever arm may enhance magnetic braking 
when $\theta=90^\circ$ and suppress it
when $\theta=0^\circ$ in the previous section.

To clarify the issue decisively,
we investigated the evolution of the angular momentum of the fluid elements
of a sphere which is selected from the initial cloud core.
Let the sphere be parameterized by the enclosed mass,
\begin{eqnarray}
M_{\rm e}(r)=\int_0^r \rho(r) d\mathbf{V}.
\end{eqnarray}
We chose a sphere with 
$M_{\rm e}(r)=5 \times 10^{-2}\msun$, i.e., the innermost
region of the entire cloud core.
Then $10^4$ SPH particles were sampled 
from this sphere at the beginning of the simulation and 
their trajectories were traced.
By this sampling procedure, the fluid elements of the simulations 
with the same $\alpha$ are the same and we can investigate 
the difference of the magnetic-braking efficiency purely caused by
misalignment.

\begin{figure*}
\includegraphics[width=40mm,angle=-90]{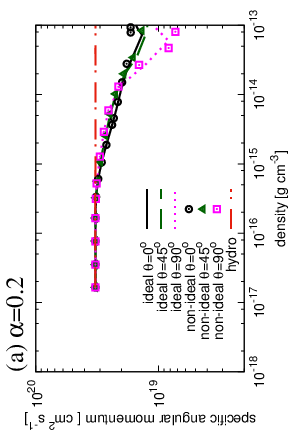}
\includegraphics[width=40mm,angle=-90]{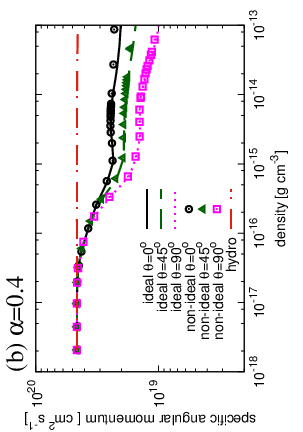}
\includegraphics[width=40mm,angle=-90]{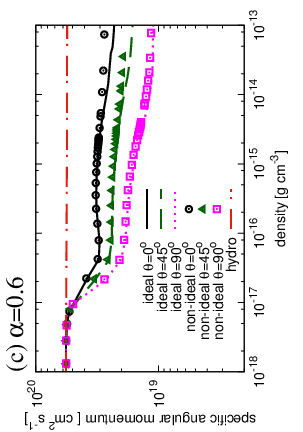}
\caption{
Temporal evolution of the mean specific angular momentum of the fluid elements 
in a sphere with $M_e(r)=5 \times 10^{-2} \msun$. 
The results of the ideal (lines) and non-ideal (symbols) MHD simulations are shown (Model I1-I9 and NI1-NI9, respectively).
The dashed-dotted lines show the results of the hydrodynamics simulations.
The horizontal axis shows the median density of the fluid elements.
The vertical axis shows the mean specific angular 
momentum of the fluid elements.
}
\label{evolution_MHD_J_sphere}
\end{figure*}

Figure \ref{evolution_MHD_J_sphere} shows 
the evolution of the mean angular momentum 
of the sphere in the isothermal collapse phase.
The results of both the ideal (lines) and non-ideal (symbols) MHD
simulations are shown.
The figure clearly indicates that the angular momentum 
becomes smaller in the simulations with
$\theta=90^\circ$ (dotted lines) than with $\theta=0^\circ$ (solid lines)
during the isothermal-collapse phase.
The initial distributions of fluid elements in each panel
are identical; hence the difference in the angular momentum is 
caused purely by the difference of the magnetic-braking efficiency that originates in
misalignment.
Therefore, we conclude that magnetic braking in the isothermal collapse phase
is more effective in the perpendicular than in the parallel configuration.

Another important finding is that, with smaller $\alpha$ values,
magnetic braking in the isothermal-collapse phase becomes weak, and 
the difference in the angular momenta between simulations
with different $\theta$ value is small.
Therefore, in the simulations with a small $\alpha$ values,
magnetic braking in the high density region plays a crucial role
in determining whether the central angular momentum of $\theta=0^\circ$ is greater 
than that of $\theta=90^\circ$.
The figure also shows that the evolution of the angular momentum is almost 
identical in the isothermal-collapse phase between the ideal and non-ideal MHD
simulations.

The temporal evolution of the angular momentum of the hydrodynamical simulations 
(more precisely, ideal MHD simulations with a very weak magnetic field of $\mu=10^{4}$)
is plotted with dashed-dotted lines.
The lines show that the angular momentum is almost constant during the isothermal-collapse 
phase. This confirms that the decrease of the angular momentum observed in
MHD simulations is indeed caused by the magnetic field.

\section{Discussion}
\label{discussion}
\subsection{Three mechanisms for the evolution of the angular momentum 
during the cloud core collapse and their characteristics}
In \S \ref{ret_3}, we showed that three mechanisms affect
the angular momentum evolution of the central region,
i.e.,
\begin{enumerate}
\item the selective accretion of the rapidly (for the core with $\theta=90^\circ$) or slowly (for the core with $\theta=0^\circ$) rotating fluid elements to the central region;
\item magnetic braking in the isothermal-collapse phase;
\item magnetic braking in the first-core (adiabatic-collapse phase) or the disk.
\end{enumerate}
We summarize the characteristics 
of these mechanisms and their dependence on the initial conditions and magnetic diffusions.

We have shown that magnetic braking in the isothermal-collapse phase
is stronger in the cores with $\theta=90^\circ$ than with $\theta=0^\circ$
by following the evolution of the angular momentum of the fluid elements.
This result is consistent with the analytic argument 
by \citet{1985A&A...142...41M} and the simulation result by \citet{2004ApJ...616..266M}.
The enhancement of magnetic braking by flared magnetic-field geometry
which was proposed as a cause for
stronger magnetic braking in the case of $\theta=0^\circ$ 
in \citet{2012A&A...543A.128J}
does not seem to be effective in the isothermal collapse phase.

\citet{2012A&A...543A.128J} pointed out that the timescale of magnetic braking  of
a core with $\theta=0^\circ$ and a flared magnetic field geometry  
becomes smaller than that with $\theta=90^\circ$.
According to their estimate, 
the magnetic-braking timescale of the core with $\theta=0^\circ$ and a field geometry
of panel (c) of figure \ref{schematic}
is smaller than that with $\theta=90^\circ$ 
 and a field geometry of panel (b) of figure \ref{schematic}
by a factor of $(R_{\rm c}/R_{\rm core}) (\ll 1)$ times,
where $R_{\rm c}$  and $R_{\rm core}$ are the radius 
of the central region and of the initial core, respectively. 
This argument seems to contradict our conclusions 
that magnetic braking during isothermal-collapse
phase tends to be strong 
in the cores with $\theta=90^\circ$. An explanation is therefore required.
In Appendix A, we compare the timescales for magnetic braking with various field geometries and
show that, even with a flared magnetic field geometry,  the timescale for magnetic braking
for the core with $\theta=0^\circ$ is not always smaller than that with $\theta=90^\circ$.
Thus, our results do not contradict the analytical estimate. Rather, the analytical estimates
have large uncertainties and even the qualitative result depends on the assumptions made.

The efficiency of magnetic braking
in the isothermal-collapse phase decreases as  $\alpha$ decreases
(figures \ref{evolution_ideal_MHD_J_center} and \ref{evolution_MHD_J_sphere}).
In this study, we parameterized the initial cloud core with 
$\alpha \equiv E_{\rm therm}/E_{\rm grav}$.
Because the initial temperature is fixed to $10$ K, 
a small $\alpha$ corresponds 
to a large gravitational energy and short free-fall timescale.
The magnetic field does not then  have
enough time to change the angular momentum of the fluid elements
during the isothermal-collapse phase, which in turn makes
the magnetic braking becomes weak in the core with small $\alpha$.

The selective accretion apparently 
weakens magnetic braking in the simulations
with $\theta=90^\circ$ and strengthens it with $\theta=0^\circ$.
The radial magnetic tension suppresses mass accretion 
from the direction perpendicular to the field line 
and selectively enhances mass accretion parallel to the
field line in the early evolution  phase.
As a result, the fluid elements with a greater initial
angular momentum selectively accrete onto the central 
region in the simulations with $\theta=90^\circ$, while
those with a smaller angular momentum selectively accrete with $\theta=0^\circ$.
The effect of selective accretion is less significant  
in a core with a smaller $\alpha$ (figure \ref{init_dist_particle}).

Selective accretion may serve as an additional mechanism for increasing
the central angular momentum in a core with
$\theta=90^\circ$ in its early evolution phase.
However, this effect is ``apparent" in the sense that 
fluid elements with a smaller angular momentum 
eventually accrete to the central region in the subsequent evolution.
In contrast, 
the disk is expected to grow in the later evolutionally phase in 
the core with $\theta=0^\circ$ because
the gas with a larger angular momentum eventually accretes to the center.
Indeed, \citet{2011PASJ...63..555M} 
reported  disk growth in the later evolutionary phase with
long-term simulations up to $10^5$ years after the protostar formation.
They suggested that the growth of the disk in the late phase 
is caused by the decrease in the magnetic-braking efficiency brought about by the
depletion of the envelope above the circumstellar disk.
However, we suggest that the delay in the accretion
of fluid elements with a large angular momentum serves as another 
mechanism for the rapid increase of the disk size
in the late phase.

Of the three mechanisms, 
magnetic braking in the high-density region (in the first core and disk) 
may depend weakly on the gravitational energy of the initial cloud core 
because the gas in the high density region
is supported by the pressure gradient force or the 
centrifugal force, and because the free-fall time of the 
initial cloud core is not the characteristic time-scale.
Owing the pressure and rotation support, 
the magnetic field has much more time than the local free-fall 
time to extract the gas angular momentum.
Because the magnetic field fans out around the central region,
the magnetic braking in the high density region of the cores
with $\theta=0^\circ$ may be enhanced by the hour-glass like
magnetic field geometry as indicated by equations (\ref{tb31}) 
and (\ref{tb32_ratio}).
This, along with non-spherical accretion, may cause the 
flipping of the central angular momentum 
between the core with $\theta=0^\circ$ 
and the core with $\theta=90^\circ$ in ideal MHD simulations of $\alpha=0.2$.

\subsection{Importance of the non-ideal effect}
The efficiency of magnetic braking in the high-density region
depends strongly on whether magnetic diffusions are taken into account.
In the ideal MHD simulations, the significant decrease
in the angular momentum is observed in the high-density region 
(figure \ref{evolution_ideal_MHD_J_center}).
Once magnetic diffusions have been taken into account, however,  
the decrease in the angular momentum becomes moderate (see, arrows (iii) in figure
\ref{evolution_resistive_MHD_J_center}).
The ambipolar and ohmic diffusion are effective in the region with $\rho\gtrsim10^{-13}\gcm$
and in $\rho\gtrsim10^{-11}\gcm$, respectively \citep{2012A&A...541A..35D,
2015MNRAS.452..278T, 2015ApJ...801..117T,
2016A&A...587A..32M,2016MNRAS.457.1037W},
and the magnetic flux is removed from the
central region. Furthermore, induction of the toroidal magnetic field 
by gas rotation is also suppressed.
As a result, the toroidal magnetic tension in the high density region
is weak in the non-ideal MHD simulations.
The simulations of disk formation with the ideal MHD simulations
may exaggerate the impact of magnetic braking 
in the high density region.
Our results show that the difference between 
the ideal and non-ideal MHD simulations is
significant and suggest that the inclusion of
magnetic diffusion is crucial for investigating
circumstellar disk formation and evolution.


The effect of the magnetic diffusions depends sensitively on the
resistivity models e.g., cosmic-ray-ionization rate, 
and the dust models \citep{1991ApJ...368..181N,2012A&A...541A..35D,
2017A&A...603A.105D,2018arXiv180303062Z}.
Their difference must therefore to borne 
in mind when comparing the previous studies
with non-ideal effects
\citep[e.g.,][]{2015ApJ...801..117T,
2015MNRAS.452..278T,2016MNRAS.457.1037W,
2016A&A...587A..32M,2016MNRAS.460.2050Z,
2018MNRAS.475.1859W,2018arXiv180108193V,
2018MNRAS.473.4868Z} quantitatively, although our results agree qualitatively 
with those earlier studies.
A thorough  systematic study of the effect of the variety of the 
resistivity model remains to be done \citep[see, however, the impact of 
the difference in cosmic ray ionization rate;][]{2018MNRAS.475.1859W,
2018MNRAS.tmp..378W}.
Furthermore, it would be worthwhile to investigate the impact of
dust growth in the disk which may happen even in the earliest 
phase of the disk evolution \citep[][]{2017ApJ...838..151T}.
We believe that the detailed study of the resistivity models is 
important subject for future work.

The present study neglects the Hall effect.
Its impact on the evolution of angular momentum in 
misaligned cloud cores is studied in detail in \citet{2017PASJ...69...95T} 
which considers a uniform cloud core with
$\alpha~0.3$ and a uniform magnetic field with $\mu=4$.
The Hall effect induces rotation in the left-handed screw direction of magnetic field
during the isothermal-collapse phase. As a result, 
the angular momentum of the 
central region (e.g., $\rho>10^{-12}\gcm$) of the core with $\theta=180^\circ$ is an order
of magnitude greater with $\theta=0^\circ$ in \citet{2017PASJ...69...95T}. 
The angular momenta of these two cases become identical
when the Hall effect is neglected.
The angular momentum of the core with $\theta=90^\circ$ has an 
intermediate value between these two cases.
\citet{2017PASJ...69...95T}, however, only investigated a single value of $\alpha$,
and it is therefore unclear how the impact of Hall effect changes in the core 
with different value of $\alpha$.
Magnetic braking in the isothermal-collapse phase 
can be expected to become more significant with a greater $\alpha$. Therefore, 
the quantitative difference between
cores with different $\theta$ values may depend on $\alpha$.
We intend to investigate this issue in future works.



\subsection{Comparison with previous studies}
\label{comp_prev}

\citet{2004ApJ...616..266M} reported that magnetic braking
is more efficient in a 
core with $\theta=90^\circ$ than with $\theta=0^\circ$
\citep[figure 12 of][]{2004ApJ...616..266M}.
Their simulations adopted the ideal MHD approximation and 
modeled the cloud core as a Bonnor-Ebert sphere
with a uniform magnetic field. 
The parameter $\alpha$ of the core was $\alpha=0.5$.
The simulations were terminated at the protostar formation epoch.
Although the density profile is different from our initial condition, 
their result is consistent with ours, obtained from ideal MHD simulations
with $\alpha=0.6$ and $\alpha=0.4$. With such large $\alpha$ values, magnetic braking in 
the isothermal-collapse phase 
plays a crucial role in the evolution of angular momentum.

\citet{2009A&A...506L..29H} and \citet{2012A&A...543A.128J} reported that 
magnetic braking is more efficient in a core with $\theta=0^\circ$ than 
with $\theta=90^\circ$, in apparent contradiction with
by \citet{2004ApJ...616..266M}.
Note, however,  that the difference in results of
\citet{2009A&A...506L..29H}, \citet{2012A&A...543A.128J} and \citet{2004ApJ...616..266M}
not contradictory but originate from different samplings 
of the parameter space, as we discuss bellow.
\citet{2009A&A...506L..29H}, \citet{2012A&A...543A.128J} also adopted the ideal MHD approximation,
and the cloud core was modeled as a gas sphere with a Plummer-like
density profile.
The magnetic field strength in \citet{2009A&A...506L..29H} was set to be proportional to 
the column density, whereas no mentions 
of the configuration of the magnetic field was given in \citet{2012A&A...543A.128J}.
Their cores have $\alpha=0.25$.
Using a barotropic equation of state in which the second collapse 
is artificially suppressed, 
\citet{2012A&A...543A.128J} followed the simulation for several thousand years after 
protostar formation.
Figure 4 in \citet{2012A&A...543A.128J} shows that the
angular momentum of the central region of the core with $\theta=0^\circ$
is smaller than that with $\theta=90^\circ$.
Although the density profile is 
different from our initial condition, 
this result is consistent with ours, obtained with
ideal MHD simulations with $\alpha=0.2$.
With this small $\alpha$ value, magnetic 
braking in the isothermal-collapse phase 
is weak and the evolution of the central angular momenta
depends more strongly on the magnetic-braking 
efficiency in the high-density region.

Indeed, the results by \citet{2012A&A...543A.128J} support our
conclusion that the evolution of the central angular momentum 
in the core with a small $\alpha$ value
is determined by magnetic braking in the high density region.
Figure 4 in \citet{2012A&A...543A.128J}
shows that the difference in
the mean angular momenta between models with different $\theta$ values is
only apparent in the high density region $n_c>10^{10} {\rm cm^{-3}}$, and that
the mean angular momentum, in the range including the low density region 
($n_c>10^{8} {\rm cm^{-3}}$),
is almost independent of $\theta$. This result suggests that 
magnetic braking in the isothermal-collapse phase does not introduce a
significant difference between the simulations with different $\theta$ values
and that the difference is caused by magnetic braking in 
the high density region.

We therefore conclude that 
the apparent discrepancy between \citet{2004ApJ...616..266M} 
and \citet{2012A&A...543A.128J} is due to the magnetic-braking 
efficiency in the isothermal collapse phase.
\citet{2004ApJ...616..266M} employed a relatively large $\alpha$ value
and significant fraction of the angular momentum is removed 
during the isothermal-collapse phase in 
their simulations with $\theta=90^\circ$.
Whether the central 
angular momenta of $\theta=0^\circ$ is greater than that of $\theta=90^\circ$ 
is determined in this phase.
This is inferred also from figure 12 of \citet{2004ApJ...616..266M},
which shows that the  difference in the central angular momenta 
is introduced in the isothermal-collapse phase.
In contrast, \citet{2009A&A...506L..29H} and \citet{2012A&A...543A.128J}
adopted a smaller $\alpha$ value than  \citet{2004ApJ...616..266M}. 
A smaller $\alpha$ value makes the magnetic braking
in the isothermal-collapse phase less significant and the difference in the 
angular momenta at the end of the isothermal collapse phase 
between models with  $\theta=0^\circ$ and $\theta=90^\circ$ becomes
small  (see, figures \ref{evolution_ideal_MHD_J_center} 
and \ref{evolution_MHD_J_sphere}).
Because magnetic braking in the high density region 
is significant in the ideal MHD simulation and is possibly stronger
in the core with $\theta=0^\circ$ than with $\theta=90^\circ$ owing
to the hour-glass like magnetic field (see Appendix \ref{analytic_arg}),
the central angular momentum in the core with 
$\theta=0^\circ$ is smaller than that with $\theta=90^\circ$.
The selective accretion may also assist that the angular momentum of $\theta=0^\circ$
become smaller than that of  $\theta=90^\circ$.

We note, however, that 
the magnetic braking in the high density region
is strongly affected by whether the magnetic diffusions are included.
Figures \ref{rhoc_JM12_ideal}, \ref{rhoc_JM12_nonideal}, \ref{evolution_ideal_MHD_J_center}, 
and \ref{evolution_resistive_MHD_J_center} show  significant differences
in the evolution of the angular momentum in the high-density region.

\citet{2016A&A...587A..32M} reported that 
the small angular misalignment of $\magB$ and $\Jang$ ($\theta<40^\circ$)
does not change the disk angular momentum noticeably.
They considered ohmic and ambipolar diffusions and 
modeled the cloud core as a uniform gas sphere with a
constant magnetic field and $\alpha=0.25$.
They adopted a barotropic equation of state
and conducted simulations 
for several thousand years after the formation of the first-core.
Figure 13 in \citet{2016A&A...587A..32M} shows that the evolution of
angular momentum of the disk is 
almost identical for $\theta=0^\circ$ 
and  $\theta=40^\circ$.
Their results are consistent with ours, shown in figure 
\ref{rhoc_JM12_nonideal} because there is only a small difference 
between the cases with $\theta=0^\circ$ and $45^\circ$.

Once magnetic diffusion is taken into account,
magnetic braking in the high density region is suppressed 
(arrows (iii) in figure \ref{evolution_resistive_MHD_J_center})
and the effect of magnetic braking in the isothermal
phase is preserved in the central region.
As a result, the central angular momentum decreases as $\theta$ increases,
independently of $\alpha$ (figure \ref{rhoc_JM12_nonideal}).
We conclude that this is the primary reason 
why a significant difference in angular momentum 
is not observed in \citet{2016A&A...587A..32M}, where
$\alpha=0.25$ and $\theta<40^\circ$.

In consideration of previous studies, our results are inconsistent with those of 
\citet{2013ApJ...774...82L}, who considered a core with uniform density
with  $\alpha=0.7$ and employed ideal MHD simulations.
They reported that circumstellar disks form more easily
in cores with $\theta=90^\circ$ than 
with $\theta=0^\circ$ even with such a large value for $\alpha$.
This contradicts our simulation results and the results of 
\citet{2004ApJ...616..266M}.

The reason for this discrepancy is unclear.
One possible reason is
the different treatment of the inner boundary.
\citet{2013ApJ...774...82L} used an open inner boundary and removed 
the gas from the system. The gas angular momentum was also removed simultaneously. 
This treatment is different from that used in other studies
\citep{2004ApJ...616..266M,2009A&A...506L..29H,
2012A&A...543A.128J,2016A&A...587A..32M} 
in which an open inner boundary was not used,
and the gas accumulated in the central region.
The first-core and the disk around it serve as a reservoir for the
angular momentum. Thus, the angular momentum also accumulates around the center.
In contrast, the use of an open inner boundary suppresses
such an accumulation of angular momentum.

Although we have reproduced almost all the qualitative results reported in previous studies,
some inconsistencies remain, as discussed above.
These may result from differences in other simulation configurations, e.g.,
boundary conditions, initial cloud cores, 
the equation of state, or numerical schemes.
We think the analysis of the evolution of the angular momentum evolution of the
fluid elements provides precise information on magnetic braking.
For future studies of this subject, we suggest 
analyzing the evolution of the angular momentum of Lagrangian fluid element
as described in this paper.

\section{Conclusions}
\label{conclusion}
Our non-ideal MHD simulations have revealed that circumstellar disks are 
 formed more easily in a core where the rotation axis is aligned with the magnetic field
direction, independently of the gravitational stability parameter $\alpha$.
This is because magnetic braking in the isothermal-collapse phase is stronger
in a core with $\theta=90^\circ$ than with $\theta=0^\circ$.
Also, magnetic diffusion suppresses magnetic braking in the first-core and circumstellar disk.
The role of magnetic diffusions has shown to be crucial, and ideal MHD simulations
may exaggerate magnetic braking in circumstellar disks.

\section *{Acknowledgments}
We thank Tomoyuki Hanawa and Tomoaki Matsumoto for fruitful discussion.
We also thank anonymous referee for helpful comments which greatly improve this paper.
We thank K. Tomida and Y. Hori for providing us with
their equation of state which was used in \citep{2013ApJ...763....6T} to us.
The computations were performed on a parallel computer, XC40/XC50 
system at Center for Computational Astrophysics of National Astronomical observatory of Japan.
This work is supported by JSPS KAKENHI grant number 17KK0096, 17K05387,
17H06360, 17H02869, 18K13581.

\appendix
\section{Analytical discussion of the magnetic braking}
\label{analytic_arg}
In this section, we derive the magnetic-braking timescales 
based on the analytical model 
of \citet{1979ApJ...230..204M,1980ApJ...237..877M} and \citet{1985A&A...142...41M}.
We particularly emphasize that a flared magnetic field geometry does not always cause the 
order reversal of the magnetic-braking timescale for the  parallel and perpendicular configuration.
\citet{1979ApJ...230..204M,1980ApJ...237..877M} showed that the magnetic-braking 
timescale $t_{\rm b}$ of a central region 
with a moment of inertia $I_{\rm c}$ can be estimated as the timescale
over which Alfv\'{e}n waves sweep through an amount of gas in the
outer envelope with a  moment of inertia $I_{\rm ext}(t_{\rm b})$ 
equals to $I_{\rm c}$.
This condition is given by
\begin{eqnarray}
\label{IeqI}
I_{\rm ext}(t_{\rm b})=I_{\rm c}.
\end{eqnarray}
This then determine the magnetic-braking timescale 
if the structures of the central region, outer envelope, 
and magnetic field are specified.

\begin{figure*}
\includegraphics[width=80mm]{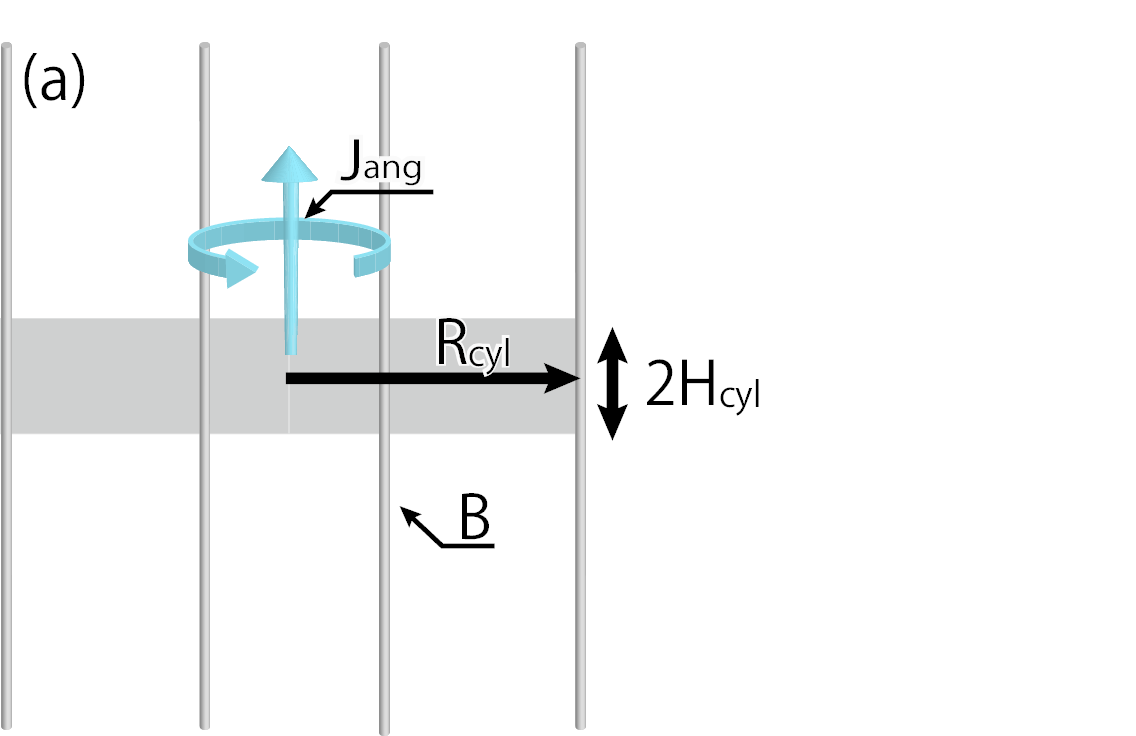}
\includegraphics[width=80mm]{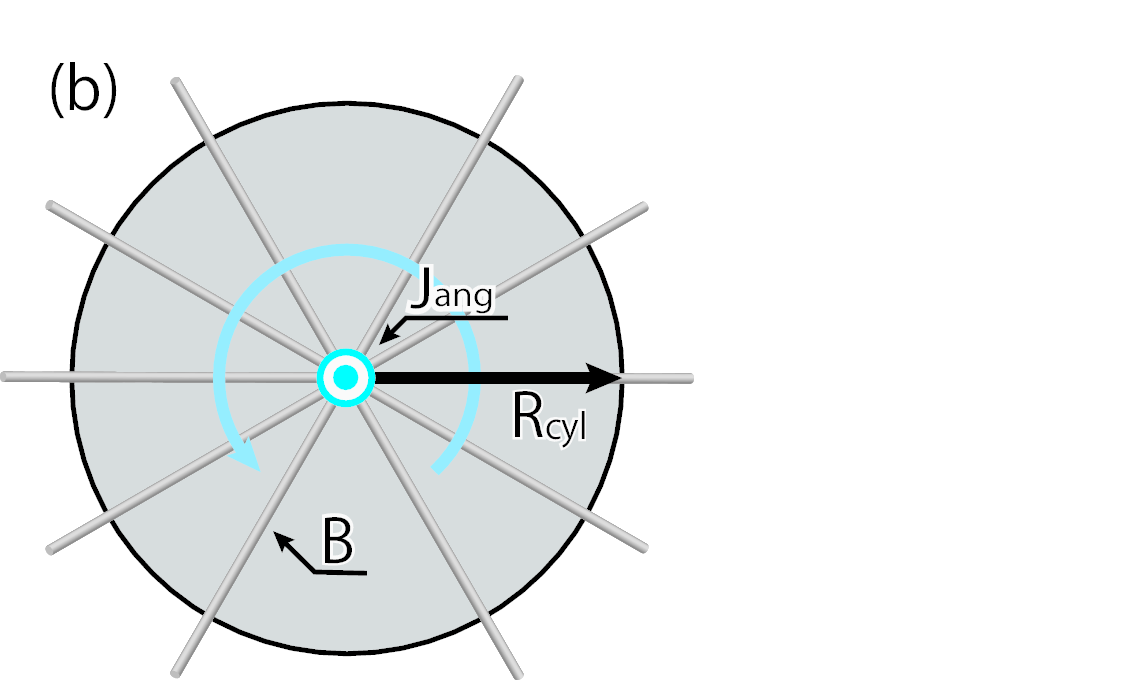}
\includegraphics[width=80mm]{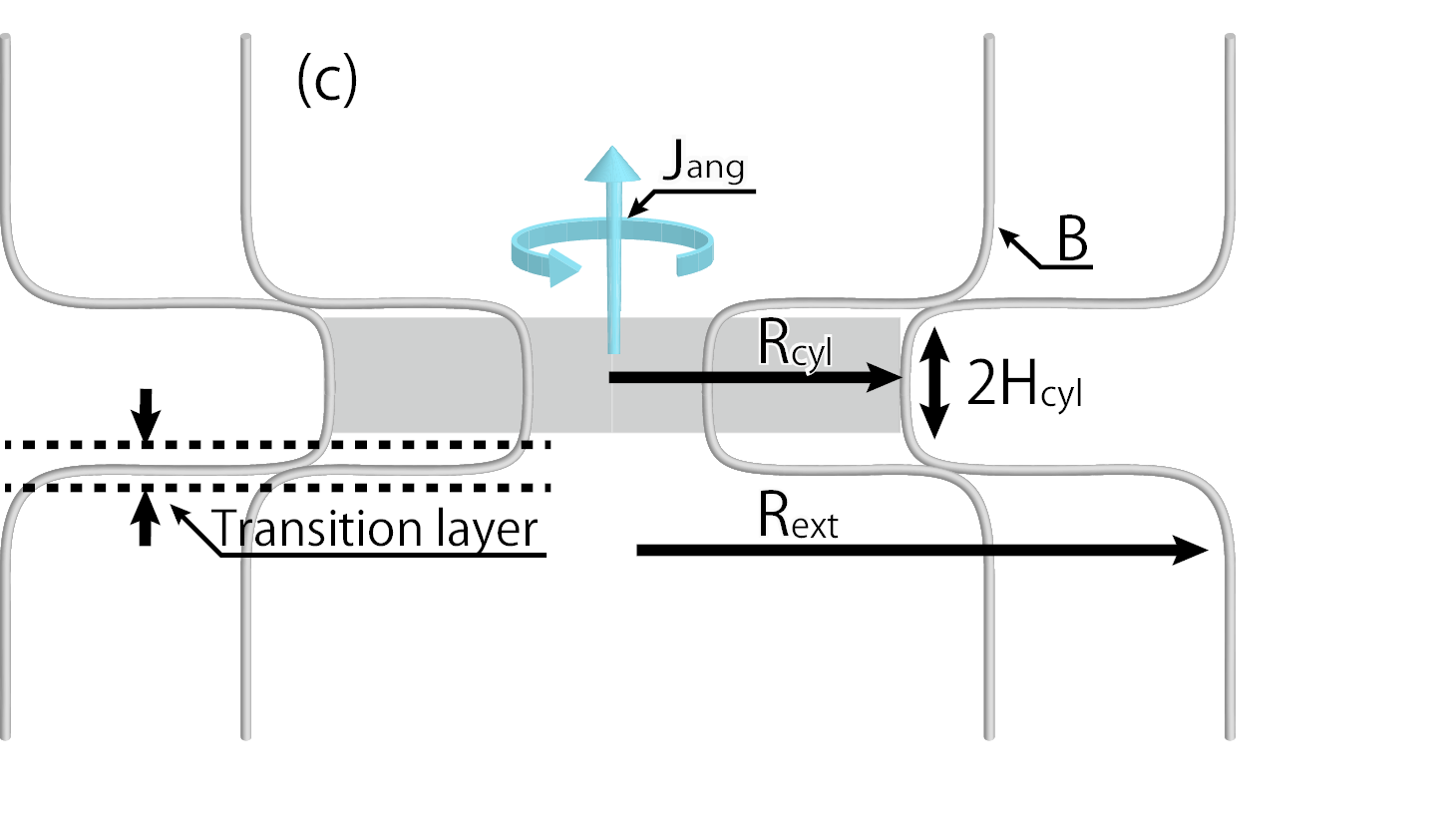}
\includegraphics[width=80mm]{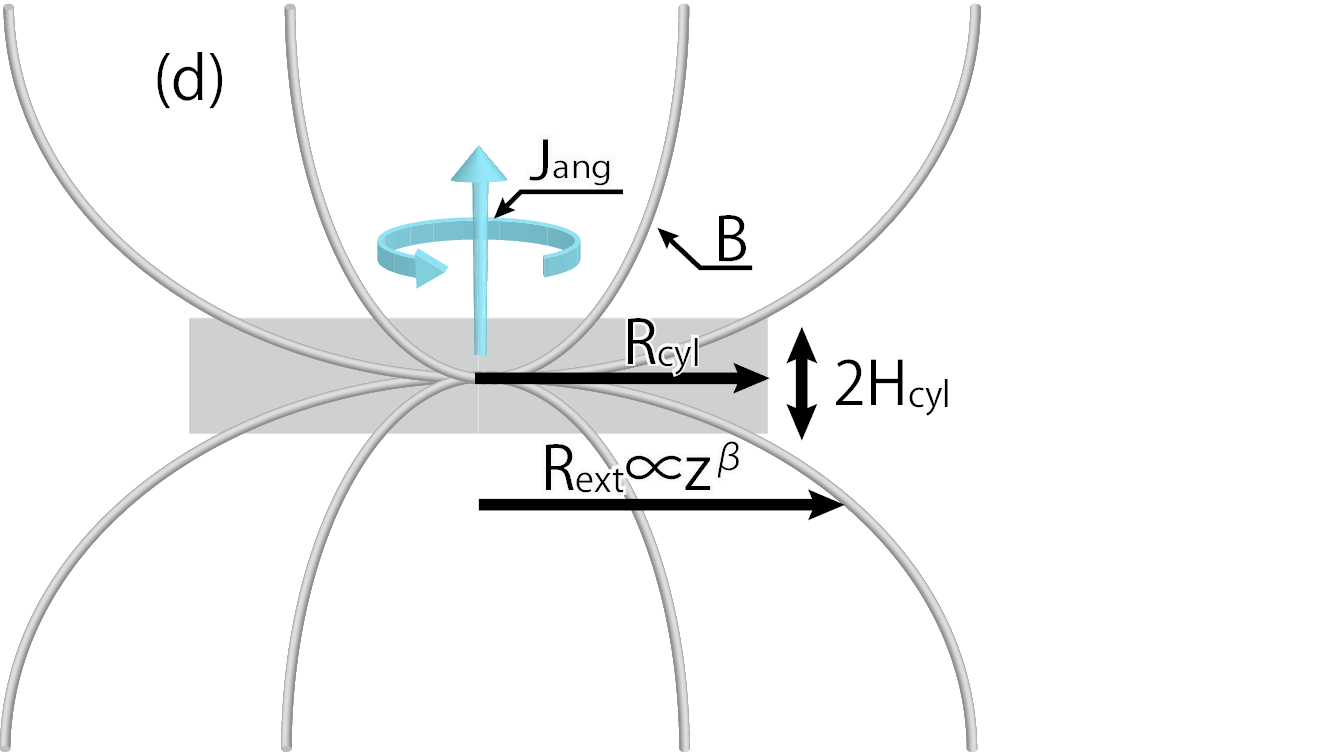}
\caption{
Schematic diagrams of the magnetic field geometries discussed 
in Appendix \ref{analytic_arg}.
Panel (a), (b), (c), and (d) show the geometry for equations
(\ref{tb11}), (\ref{tb_perp1}), (\ref{tb2}), and (\ref{tb31}),
respectively.
}
\label{schematic}
\end{figure*}

The magnetic-braking timescale for a
simple parallel configuration ($\Jang \parallel \magB$ 
or $\theta=0^\circ$) can be calculated as follows. The central collapsing region
is modeled as a uniform cylinder with a density $\rho_{\rm cyl}$,
radius $R_{\rm cyl}$, and height $2 H_{\rm cyl}$, where a uniform
magnetic field runs parallel to the rotation axis and
the density of the outer envelope  
$\rho_{\rm ext}$ is assumed to be constant (panel (a) in figure \ref{schematic}).
The moments of inertia of the central {\rm cylinder} and outer envelope are given by
$I_{\rm c}=\pi \rho_{\rm cyl} R_{\rm cyl}^4 H_{\rm cyl}$ and
$I_{\rm ext}(t_{\rm b})=\pi \rho_{\rm ext} R_{\rm cyl}^4 v_{\rm A} t_{\rm b}$,
respectively, where $v_{\rm A}$ denotes the Alfv\'{e}n velocity
in the outer envelope and $v_{\rm A} t_{\rm b}$ corresponds to the
height of the swept outer envelope.
By solving equation (\ref{IeqI}),
$t_{\rm b}$ is calculated to be 
\begin{eqnarray}
\label{tb11}
t_{\rm b, \parallel}=\frac{\rho_{\rm cyl}}{\rho_{\rm ext}}\frac{H_{\rm cyl}}{v_{\rm A}},
\end{eqnarray}
\citep{1985A&A...142...41M}.
Using the mass 
$M=2\pi\rho_{\rm cyl} R_{\rm cyl}^2 H_{\rm cyl}$ and the magnetic 
flux $\Phi=\pi R^2_{\rm cyl} B$
of the cylinder, we can reduce $t_{\rm b, \parallel}$ to
\begin{eqnarray}
\label{tb12}
t_{\rm b,\parallel} = \left( \frac{\pi}{\rho_{\rm ext}} \right)^{1/2} \frac{M}{\Phi}.
\end{eqnarray}
This shows that the magnetic-braking timescale 
in the simple geometry depends only on
the mass-to-flux ratio of the central region
and the density of the outer envelope.

Magnetic-braking timescale in a
simple geometry with a perpendicular configuration ($\Jang \perp \magB$ 
or $\theta=90^\circ$) is calculated as follows.
The central collapsing region
is modeled in the same way as with the parallel configuration, 
using the parameters $\rho_{\rm cyl}$, $R_{\rm cyl}$, and $2 H_{\rm cyl}$.
The magnetic field is assumed to be perpendicular to the rotation axis and axisymmetrical
(panel (b) in figure \ref{schematic}) where $B(r) \propto r^{-1}$.
The density of the outer envelope  $\rho_{\rm ext}$ is again 
assumed to be constant (panel (b) of figure \ref{schematic}).

The moments of inertia of the central {\rm cylinder} and outer envelope are given by
$I_{\rm c}=4 \pi \rho_{\rm cyl} H_{\rm cyl} R_{\rm cyl}^4 $ and
$I_{\rm ext}=4 \pi \rho_{\rm ext} H_{\rm cyl} (R_{\rm ext}^4 - R_{\rm cyl}^4) $.
Using the condition  $I_{\rm c}=I_{\rm ext}$,
the magnetic-braking timescale for the perpendicular configuration is
calculated to be
\begin{eqnarray}
\label{tb_perp1}
t_{\rm b,\perp} =\int^{R_{\rm ext}}_{R_{\rm cyl}}\frac{dr}{v_{\rm A}}=\frac{1}{2} \left(\left( \frac{\rho_{\rm cyl}}{\rho_{\rm ext}} 
+1\right)^{1/2} -1\right)\frac{R_{\rm cyl}}{v_{\rm A}(R_{\rm cyl})} \nonumber \\
  \sim \frac{1}{2} \left( \frac{\rho_{\rm cyl}}{\rho_{\rm ext}} 
\right)^{1/2}\frac{R_{\rm cyl}}{v_{\rm A}(R_{\rm cyl})},
\end{eqnarray}
\citep{1985A&A...142...41M}, where we assumed $\rho_{\rm cyl}/\rho_{\rm ext}\gg 1$.
$t_{\rm b,\perp}$ is reduced to
\begin{eqnarray}
\label{tb_perp2}
t_{\rm b,\perp} \sim 2 \left( \frac{\pi}{\rho_{\rm cyl}} \right)^\frac{1}{2}\frac{M}{\Phi}.
\end{eqnarray}
Note that this formula is likely to overestimate the timescale 
(or to underestimate the magnetic-braking efficiency), because, in
the magnetic field configuration, the Alfv\'{e}n velocity decreases as 
$v_A \propto r^{-1}$ and a long lever arm cannot be obtained.
If the magnetic field has the anisotropic configuration, the 
Alfv\'{e}n wave can propagate further and the longer lever arm 
decreases the braking timescale.

The ratio of the magnetic-braking 
timescale for the perpendicular and the parallel configurations is
\begin{eqnarray}
\label{ratio1}
\frac{t_{\rm b,\perp}}{t_{\rm b, \parallel}} \sim \left( \frac{\rho_{\rm ext}}{\rho_{\rm cyl}} \right)^{1/2}.
\end{eqnarray}
Thus the timescale for
perpendicular configuration is shorter than that for the
parallel one because $\rho_{\rm cyl} \gg \rho_{\rm ext}$ and 
the magnetic braking in the former is stronger than in the later.
This is the straightforward conclusion derived for the simplest geometry cases.

However, 
the simple geometry of the  parallel configuration may be inappropriate for 
a model of a collapsing cloud core.
The magnetic field is suggested to have an hour-glass-like shape
as a result of radial 
dragging according to the theoretical work of \citep[e.g.,][]{2002ApJ...575..306T,2003ApJ...599..363A,
2010MNRAS.408..322K,2015MNRAS.452..278T,2018arXiv180108193V}, and
as has been confirmed observationally in YSOs \citep{2006Sci...313..812G}.
In the hourglass configuration,
the magnetic field fans out 
vertically, which is inconsistent
with the simple geometry of the parallel configuration.

\citet{1985A&A...142...41M} derived the magnetic-braking timescale
in which the magnetic flux tube expands quickly
in an infinitely thin transition layer
to a radius $R_{\rm ext}$ just above the central cylinder
(panel (c) in figure \ref{schematic}).
If we ignore the moment of inertia of the transition layer,
$I_{\rm ext}(t_{\rm b})$ is given by
\begin{eqnarray}
\label{eq_ext2}
I_{\rm ext}(t_{\rm b})=\pi \rho_{\rm ext} R_{\rm ext}^4 v_{\rm A} t_{\rm b}.
\end{eqnarray}
Using $I_{\rm c}=\pi \rho_{\rm cyl} R_{\rm cyl}^4 H_{\rm cyl}$
and equation (\ref{eq_ext2}), we obtain the 
magnetic-braking timescale of the disk
with an hour-glass magnetic field geometry  \citep{1985A&A...142...41M} as
\begin{eqnarray}
\label{tb2}
t_{\rm b,f}=\left( \frac{\pi}{\rho_{\rm ext}} \right)^{1/2} \left(\frac{M}{\Phi}\right)
\left(\frac{R_{\rm cyl}}{R_{\rm ext}}\right)^2,
\end{eqnarray}
where
\begin{eqnarray}
\label{b_change}
B_{\rm ext}=\left(\frac{R_{\rm cyl}}{R_{\rm ext}} \right)^{2}B_{\rm cyl},
\end{eqnarray}
is assumed because of the conservation of magnetic flux.
According to equation (\ref{tb2}), the magnetic-braking timescale can be shorter
than $t_{\rm b,\parallel}$ in equation (\ref{tb12}) because of the factor $(R_{\rm cyl}/R_{\rm ext})^2 (< 1)$.

\citet{2012A&A...543A.128J} argued that the 
magnetic-braking timescale becomes smaller in the parallel 
than perpendicular configuration
by virtue of  equation (\ref{tb2}).
They assumed $R_{\rm ext}=R_{\rm core}$,
where $R_{\rm core}$ is the initial core radius, and
$\rho(r)\propto r^{-2}$. Then, assuming that the  mean 
external density and the central density can be approximated in terms of
the volume-averaged means as, respectively,  $\rho_{\rm cyl} \sim \rho(R_{\rm cyl})$ 
and $\rho_{\rm ext} \sim \rho(R_{\rm ext})$,
the ratio of the timescales is estimated to be
\begin{eqnarray}
\label{ratio1}
\frac{t_{\rm b,\perp}}{t_{\rm b,f}} \sim \frac{R_{\rm core}}{R_{\rm cyl}}.
\end{eqnarray}
Using this relation, they concluded that the braking timescale in
flared parallel configuration is shorter than in the
perpendicular configuration because $R_{\rm cyl}<R_{\rm core}$.

However, it is uncertain whether the assumption of 
an infinitely thin magnetic fluxtube expansion layer (transition layer)
is valid in realistic situations,
given that such a layer causes a strong dependence of the braking timescale on 
the ratio of the radii $t_{\rm b,f} \propto (R_{\rm cyl}/R_{\rm ext})^2$.
This assumption corresponds to discontinuous decrease 
of poloidal magnetic field strength (equation (\ref{b_change}))
at the vertical height $z=H_{\rm cyl}$.
The poloidal magnetic field in a collapsing cloud core, however,  
may change in a power-law fashion. 
Furthermore, the analytic estimate ignores the propagation time of 
the Alfv\'{e}n wave in the expansion layer, 
which would not be negligible when $R_{\rm cyl} \ll R_{\rm ext}$.

We should also note that validity of the assumption of $R_{\rm ext}=R_{\rm core}$ 
is unclear,
because $R_{\rm ext}$ is the radius of the external flux tube and
is not related to that of the core $R_{\rm core}$, and similarly because
the radius in the cylindrical coordinates (or distance from the vertical axis)
is not related to the radius in the spherical coordinates (or distance from the center).
$R_{\rm ext}$ can be $R_{\rm ext} \ll R_{\rm core}$ 
even at the vertical height of $z=R_{\rm core}$ 
unless all magnetic flux of the core is
concentrated in the central region which
is unlikely in the early phase of disk formation.
Therefore, equation (\ref{tb2}) should be regarded as 
the lower limit of the magnetic-braking timescale 
in the parallel configuration.

To demonstrate clearly that the discontinuous expansion of magnetic flux-tubes
and the assumptions described above
lead to the flip of the magnetic-braking timescale,
we calculate the magnetic-braking 
timescale with magnetic flux-tubes which expand in a power-law fashion.
Our result will show that whether the timescale becomes shorter or not depends on
the power-law index and external density in this case.

We consider the configuration, where the radius of the magnetic fluxtube obeys
\begin{eqnarray}
R_{\rm ext}(z)=R_{\rm cyl} \left( \frac{z}{H} \right)^\beta,
\end{eqnarray}
(see panel (d) in figure \ref{schematic}).
We consider  the case of $\beta>0$ only, i.e., where the flux-tube
is expanding.
The magnetic field strength is approximated to
\begin{eqnarray}
B(z) \sim B_{\rm cyl} \frac{R_{\rm cyl}^2}{R_{\rm ext}(z)^2} = B_{\rm cyl} 
\left(\frac{z}{H}\right)^{-2\beta}.
\end{eqnarray}
where the vertical magnetic field is assumed to be constant 
for a given $z$ and
the wavefront of the Alfv\'{e}n wave is assumed to be perpendicular to the vertical axis for simplicity.
The time for which the Alfv\'{e}n wave reaches $z$ is calculated to be
\begin{eqnarray}
t(z)=\int^{z}_{0} \frac{dz}{v_A(z)}.
\end{eqnarray}
where the interval of integration is approximated as $[0:z]$ 
from the exact formula of $[H_{\rm cyl}:z]$. 
Note that this simplification is also implicit
in equation (\ref{tb_perp1}).
The integration yields a solution for $z$ as a function of time as
\begin{eqnarray}
z(t)=\frac{(1+2\beta) H^{2\beta} B_{\rm cyl}}{(4 \pi \rho_{\rm ext})^{1/2}}t .
\end{eqnarray}
Then the moment of inertia of the outer region as a function of time 
is calculated to be
\begin{eqnarray}
I_{\rm ext}(t)=\int^{t}_{0} \rho_{\rm ext} R_{\rm ext}(z(t))^4 v_A(z(t)) dt,
\end{eqnarray}
where the interval of integration is again approximated to $[0:t]$, 
not to $[t_0:t]$ where $t_0$ is defined by $z(t_0)=H_0$.
By using equation (\ref{IeqI}), we finally obtain the 
magnetic-braking timescale in a gradually expanding magnetic field 
configuration as
\begin{eqnarray}
\label{tb31}
t_{\rm b, power}=\frac{1}{1+2\beta}[(1+4\beta)
(\frac{\rho_{\rm cyl}}{\rho_{\rm ext}})]^{\frac{1+2\beta}{1+4\beta}}\frac{H_{\rm cyl}}{v_A(H_{\rm cyl})}.
\end{eqnarray}
This can be directly compared with equations (\ref{tb11}) and (\ref{tb_perp1}).
Using the mass to flux ratio, equation (\ref{tb31}) 
is expressed also as
\begin{eqnarray}
\label{tb32}
t_{\rm b, power}= \frac{(1+4\beta)^{\frac{1+2\beta}{1+4\beta}}}{1+2\beta}
(\frac{\rho_{\rm cyl}}{\rho_{\rm ext}})^{\frac{-2\beta}{1+4\beta}}\left(\frac{\pi}{\rho_{\rm ext}}\right)^{1/2} \frac{M}{\Phi}.
\end{eqnarray}
When $\beta=0$, equations (\ref{tb31}) and (\ref{tb32}) 
reduce to equations (\ref{tb11}) and  (\ref{tb12}), respectively.
Also, $t_{\rm b, power}$ satisfies $t_{\rm b, power}<t_{\rm b,\parallel}$ because 
$(\rho_{\rm cyl}/\rho_{\rm ext})^{-2\beta/(1+4\beta)}<1$.
The magnetic-braking timescale 
becomes shorter because of the  expansion of the flux-tube.

The ratio of  $t_{\rm b, perp}$ to $t_{\rm b,power}$  is given by
\begin{eqnarray}
\label{tb32_ratio}
\frac{t_{\rm b,\perp}}{t_{\rm b, power}}=
\frac{2+4\beta}{(1+4\beta)^{\frac{1+2\beta}{1+4\beta}}}
\left( \frac{\rho_{\rm ext}}{\rho_{\rm cyl}}\right)^{\frac{1}{2+8\beta}}.
\end{eqnarray}
This indicates that $t_{\rm b, power}$ is not always 
smaller than $t_{\rm b,\perp}$,
and that whether $\frac{t_{\rm b,\perp}}{t_{\rm b, power}}$ is smaller or
greater than unity
depends on $\beta$ and on the density ratio.
As $\beta$ increases, the contribution of the 
density ratio diminishes quickly owing to the strong
dependence of the power index on $\beta$ in the term
$\left( \frac{\rho_{\rm cyl}}{\rho_{\rm ext}}\right)^{\frac{1}{2+8\beta}}$.
The numerical factor 
$\frac{2+4\beta}{(1+4\beta)^{\frac{1+2\beta}{1+4\beta}}}$ 
is greater than unity for $\beta>0$ and  $t_{\rm b,\perp}$ can become
greater than $t_{\rm b, power}$.
However, the numerical factor depends on
our assumptions and has a large uncertainty. 
Therefore, we cannot directly conclude the whether 
the magnetic-braking timescales of parallel 
configuration is longer or shorter than
that of perpendicular configuration on the basis of the analytic estimate.


Although the above analytic discussion
has provided physical insight into magnetic-braking,
the modeling has involved several assumptions and simplifications.
that can affect even the qualitative results.
Multi-dimensional simulations are therefore essential to achieve a
conclusive understanding of
the effect of magnetic-braking in collapsing cloud cores.

\bibliography{article}

\end{document}